\documentclass[preprint]{iucr}              

\usepackage{graphicx}
\usepackage{amsmath}
\usepackage{amssymb}
\usepackage{units}
\usepackage{braket}
\usepackage{xcolor} 
\usepackage{makecell}
\usepackage{array}
\newcolumntype{K}[1]{>{\centering\arraybackslash}p{#1}}

\begin{document}                  



\title{Review of honeycomb-based Kitaev materials with zigzag magnetic ordering}


\cauthor[]{V. Ovidiu}{Garlea}{garleao@ornl.gov}{address if different from \aff}
\author[]{Colin L.}{Sarkis}

\aff{Neutron Scattering Division, Oak Ridge National Laboratory, Oak Ridge, TN 37831, \country{USA}}







\maketitle                        

%

\begin{abstract}
The search for a Kitaev quantum spin liquid in crystalline magnetic materials has fueled intense interest in the two-dimensional honeycomb systems. Many promising candidate Kitaev systems are characterized by a long range ordered magnetic structure with an antiferromagnetic zigzag-type order, where the static moments form alternating ferromagnetic chains. Recent experiments on high-quality single crystals uncovered the existence of intriguing multi-$\mathbf{k}$ magnetic structures, which evolved from zigzag structures. Those discoveries have sparked new theoretical developments and amplified interest in these materials. We present an overview of the honeycomb materials known to display this type of magnetic structure and provide detailed crystallographic information for the possible single and multi-$\mathbf{k}$ variants.
\end{abstract}


\section{Introduction}

\begin{figure}
\includegraphics[width=3.2in]{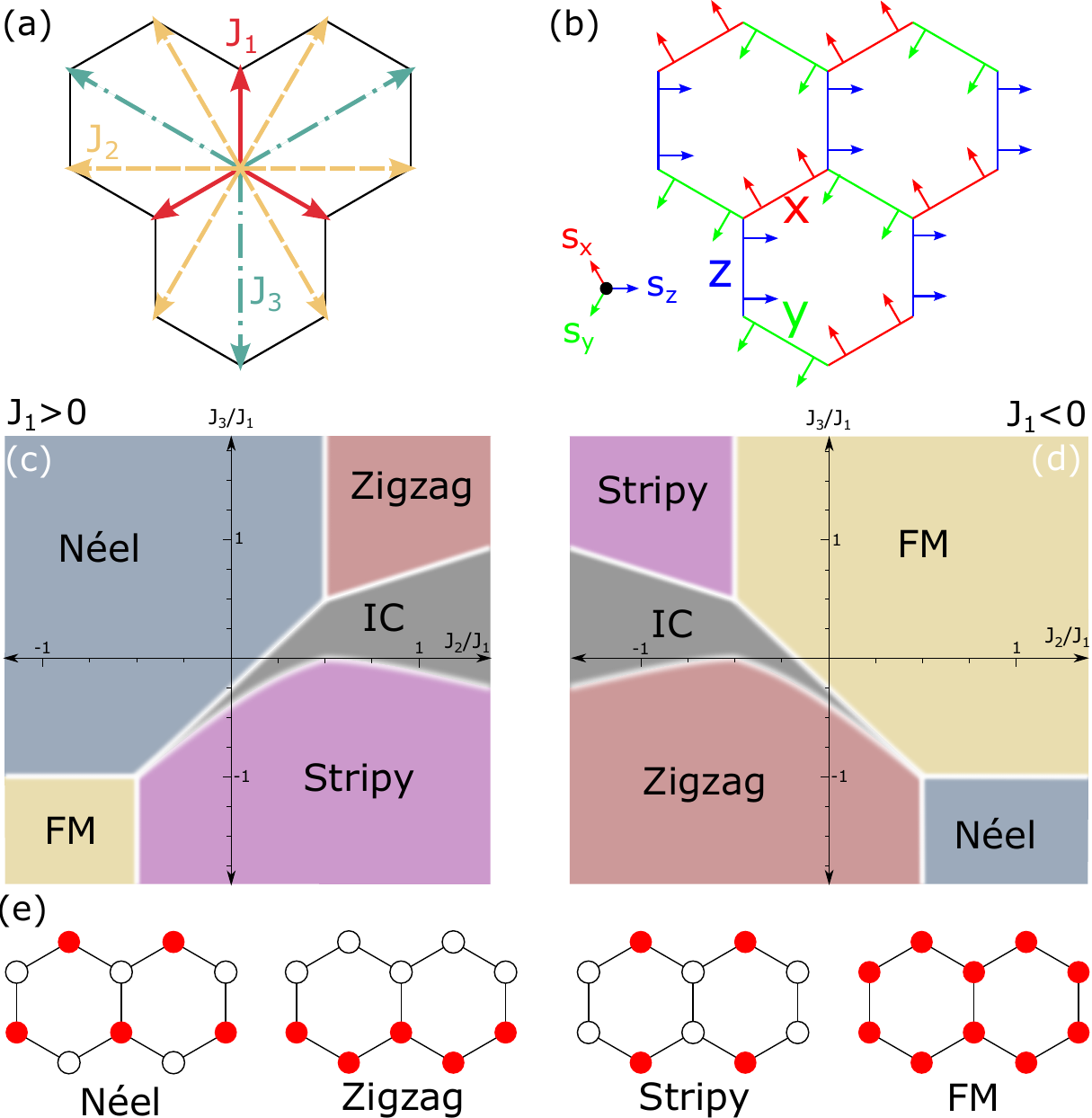}
\caption{(a) Near neighbor connectivity on the 2D honeycomb lattice. Here first nearest neighbor interactions ($J_1$) are given by solid red arrows, second nearest neighbor interactions ($J_2$) are given by gold dashed lines, and third nearest neighbors ($J_3$) are shown by teal dot-dashed lines. (b) Bond dependent nearest-neighbor Ising interactions of the Kitaev model (for FM Kitaev), showing strong exchange driven frustration between three incompatible orthogonal spin axes. $J_1$-$J_2$-$J_3$ Heisenberg classical phase diagrams of the honeycomb lattice for (c) AFM and (d) FM nearest neighbor (adapted from Fouet \emph{et al.}, 2001). The two phase diagrams are very similar and show six possible ground state configurations: one co-linear FM, three co-linear AFM states, a N\'eel type, a stripy, and a zigzag, and two incommensurate orders (presented here together as one IC phase). (e) Visual depiction of the four colinear magnetic orders described above. Red and open white circles represent oppositely aligned moments.}
\label{fig:diag} 
\end{figure}

Honeycomb materials have been the subject of much interest over the last few decades for their unique properties and promise in a wide range of applications (Seepersad \emph{et al.}, 2004; Zhang \emph{et al.}, 2015; Qi \emph{et al.}, 2021). Magnetic honeycomb materials, where magnetic ions decorate the corners of a 2D honeycomb network have been investigated in the search for a quantum spin liquid (QSL)- an exotic phase of matter where strong quantum fluctuations disrupt formation of long range order while preserving a highly correlated entangled ground state (Balents, 2010; Savary \emph{et al.}, 2016; Broholm \emph{et al.}, 2020). The closely related triangular lattice system has remained the paradigmatic example of geometrical frustration-driven QSL, ever since it was first proposed by P.W. Anderson (1973). Likewise, honeycomb systems also show great promise toward the realization of a QSL, aided both by a reduced nearest neighbor connectivity (3 in honeycomb vs 6 in triangular) of the magnetic ions within the layers (Fig. \ref{fig:diag}a), as well as a pronounced 2D nature of the magnetism, which emerges from the well separated layers.


In the decoupled 2D limit, a single honeycomb layer's magnetic ground states were explored extensively by Fouet \emph{et al.} (2001), in context of a Heisenberg Hamiltonian of the form:
\begin{equation}
    H = J_{1}\sum_{<ij>_1}  \vec{S}_{i} \cdot \vec{S}_{j} + J_{2}\sum_{<ij>_2}  \vec{S}_{i} \cdot \vec{S}_{j} + J_{3}\sum_{<ij>_3}  \vec{S}_{i} \cdot \vec{S}_{j}
\\
\end{equation}
where $J_1$ represents the strength of the nearest neighbor (NN) bond, $J_2$ the second NN bond, $J_3$ the third NN bond, and the brackets, $\braket{ij}_{n}$, indicate the sum be carried over all $n^{th}$ neighbor bonds. For both an AFM NN ($J_1 > 0$) and FM NN ($J_1 < 0$), Fouet \emph{et al.} fully characterized the exchange-space phase diagrams in the classical limit and found six distinct magnetic ground states shown in Fig.~\ref{fig:diag}c-d. These are characterized by one FM order, two different incommensurate (IC) orders, and three AFM orders: a standard N\'eel state, a stripy state, and a zigzag state where FM chains are coupled antiferromagnetically. The four colinear phases are shown in Fig.~\ref{fig:diag}e for visualization.
Fouet \emph{et al.} (2001) also explored the impact of quantum fluctuations on phase diagrams, and showed that while the colinear classical ground states persist, the boundaries can undergo significant shifts. Moreover, they reported evidence of unconventional states, especially in proximity to phase boundaries. Further investigation of the quantum phase diagram by the community showed the quantum fluctuations can stabilize many unconventional phases including valence bond crystal phases, quantum dimer phases, and quantum spin liquid phases (Moessner \emph{et al.}, 2001; Sindzingre \emph{et al.}, 2003; Zhang \emph{et al.}, 2013).

While the isotropic Heisenberg interactions can accurately describe the Hamiltonians of many magnetic systems, several honeycomb materials feature either an easy plane or easy axis anisotropy, which slightly modifies the Hamiltonian above to be an XXZ type Hamiltonain of the form:
\begin{equation}
    H = \sum_{n=1}^3 J_{n}\sum_{<ij>_n}  \left[ S_{i}^{x} S_{j}^{x} +  S_{i}^{y} S_{j}^{y} +  \lambda S_{i}^{z} S_{j}^{z}\right]
\end{equation}
Where now $\lambda$ corresponds to the anisotropy in the system, either being constrained to $\lambda \in \{0,1\}$ for an easy plane type exchange anisotropy or  $\lambda > 1$ for an easy axis anisotropy, (with $\lambda=1$ being the Heisenberg model described above). While there have been investigations of the $J_1$-$J_2$ XXZ (Li \emph{et al.}, 2014), the zigzag phase cannot be stabilized by $J_1$-$J_2$ model and thus more recent theoretical work has shifted toward a $J_1$-$J_3$ XXZ model. Density matrix renormalization group (DMRG) approaches show a similar phase diagram to the isotropic case described by Fouet \emph{et al.}  where the zigzag is stabilized across a wide parameter range for a FM nearest neighbor and AFM third neighbor interaction (Jiang \emph{et al.}, 2023; Bose \emph{et al.}, 2023).

A separate mechanism of exchange frustration was explored by Alexei Kitaev (2006) using a toy model of a $S$ = 1/2 bond dependent Ising Hamiltonian on a honeycomb lattice, where each nearest neighbor bond uniquely couples a separate spin component (depicted in Fig.~\ref{fig:diag}b for a FM case). Via a transform into a Majorana framework, Kitaev was able to show this model is exactly solvable and contains a QSL ground state, either a gapless QSL phase, if all three bond-dependent exchanges were similar in strength, or gapped in the case of one exchange being dominant over the other two. More interestingly Kitaev showed that for the gapless case, application of a perturbative magnetic field was able to realize a state with non-Abelian anyonic statistics which was previous discussed as a possible framework for the development of topological quantum computation (Bravyi \emph{et al.}, 2002; Freedman \emph{et al.}, 2002; Kitaev, 2003).

While the search for Kitaev materials represent an important focus in the search for fault tolerant topological quantum computing, there already exist multiple comprehensive review articles covering Kitaev physics and materials (see for instance: Winter \emph{et al.}, 2017; Hermanns \emph{et al.}, 2018; Takagi \emph{et al.}, 2019; Trebst \&  Hickey, 2022). In this work, rather than focusing on the QSL ground state of the pure Kitaev model, we instead focus on the mechanisms for realizing the commonly observed zigzag magnetic structure in honeycomb materials, which many of the promising Kitaev candidate materials have been found to exhibit. 

The extension of Kitaev's model to real materials was made later by Jackeli and Khaliullin (2009) centered on 5$d^5$ Ir$^{4+}$ (and later 4$d^5$ Ru$^{3+}$) with a combination of spin-orbit coupling, octahedral crystal field effects, and edge sharing geometry of neighboring octahedral clusters. For an ideal octahedral coordination the three orthogonal spin axes lie canted out of the honeycomb planes ($\alpha = arcsin(1/\sqrt{3}) \sim 35.26^\circ$), naturally lying perpendicular to the three magnetic-ligand-magnetic bonding planes. In this limit, non-bond dependent interactions cancel, leaving only the Kitaev term surviving. 

The detection of zigzag magnetic order in the primary candidate materials, Na$_2$IrO$_3$ and $\alpha$-RuCl$_3$, ultimately led to the development of a more comprehensive Kitaev model that respects symmetry and   accounts for the stabilization of order in these materials:
\begin{equation}
\begin{split}
    H = \sum_{n=1}^3 \sum_{<ij>_n} J_{n} \vec{S}_{i} \cdot \vec{S}_{j} + \sum_{<ij>} \sum_{\alpha,\beta,\gamma} [ K^{\gamma} S_{i}^{\gamma} S_{j}^{\gamma} \\
    + \Gamma \left( S_{i}^{\alpha} S_{j}^{\beta} + S_{i}^{\beta} S_{j}^{\alpha}  \right) \\ 
    + \Gamma' \left( S_{i}^{\alpha} S_{j}^{\gamma}  + S_{i}^{\gamma} S_{j}^{\alpha} + S_{i}^{\beta} S_{j}^{\gamma} + S_{i}^{\gamma} S_{j}^{\beta} \right) ]
\end{split}
\end{equation}
where alongside the Kitaev term, $K^{\gamma}$, there exist nearest neighbor off-diagonal terms $\Gamma$ and $\Gamma'$, as well as Heisenberg interactions up the third nearest neighbor. Here $n$ in the first sum indicates the $n^{th}$ rank nearest neighbor, and $\alpha$,$\beta$,$\gamma$ sum over the three bonds directions dictated by the local x,y,z coordinates. In the applicable isotropic Kitaev limit, the Kitaev term reduces to $|K|^\gamma$ = $|K|^\alpha$ = $|K|^\beta$ = $K$.

Shortly after the proposal by Jackeli and Khaluilin, multiple extensions of realizing Kitaev materials were proposed in relatively low spin orbit coupled 3$d$ materials, holding promise in opening up the relatively scarce materials landscape (Liu \&  Khaliullin, 2018; Stavropoulos \emph{et al.}, 2019; Liu \emph{et al.}, 2020; Motome \emph{et al.}, 2020). The primary subject of the 3$d$ materials have focused on cobaltates and nickelates relying on the high spin ($S$ = 3/2) Co$^{2+}$ and ($S$ = 1) Ni$^{2+}$. Most notably for cobalt in the presence of an intermediate strength octahedral crystal field, Co$^{2+}$ has shown a spin orbit coupled Kramers ground state doublet in multiple materials thanks to its relatively unquenched orbital momentum. This mechanism has made many cobalt materials hallmark examples of $S$ = 1/2 pseudospin materials (Holden \emph{et al.}, 1971; Buyers \emph{et al.}, 1971; Goff \emph{et al.}, 1995; Ringler \emph{et al.}, 2022).

While initial experimental studies reported a dominant Kitaev exchange in various cobaltates, more recent theoretical approaches have indicated that the presence of multiple hopping channels and low spin-orbit coupling typically results in a suppression of Kitaev exchange, leaving the cobaltates more at the mercy of fine tuning of local environments (Das \emph{et al.}, 2021; Winter \emph{et al.}, 2022; Maksimov \emph{et al.}, 2022; Liu \& Kee, 2023). Many of these reports also highlight a large third nearest neighbor interaction ($J_3$), due to exchange pathways through neighboring ligand complexes.

Of the multiple ground state configurations found, most candidate Kitaev materials have shown either an incommensurate spin spiral state or the colinear zigzag AFM phase. Stabilization of the zigzag phase can be realized either through combination of a FM $J_1$ and AFM $J_3$, or by dominant Kitaev with subdominant off-diagonal nearest neighbor ($\Gamma$, $\Gamma'$). While these two opposing mechanisms for development of zigzag order have been the subject of much debate in the search for Kitaev materials, the prevalence of the unusual zigzag order in honeycomb magnets is clear. Table \ref{zizag_table} highlights the stability of the zigzag structure which remains the ground state configuration across a wide range of crystal space groups ($C2/m$, $C2/c$, $P6_322$, $P6_3/mcm$, $R\bar{3}$, etc) and magnetic ion species (Co$^{2+}$, Ni$^{2+}$, Ru$^{3+}$, Ir$^{4+}$ ).

While quantum and thermal fluctuations often stabilize colinear structures (Price \emph{et al.}, 2013; Rau \emph{et al.}, 2018), more exotic non-colinear structures can be stabilized by interactions beyond standard bilinear exchange (Korshunov, 1993; Kr\"uger \emph{et al.}, 2023). In the honeycomb material Na$_2$Co$_2$TeO$_6$, Kr\"uger \emph{et al.}, (2023) demonstrated that ring exchange stabilizes a non-collinear $C_3$ symmetric triple-$\textbf{k}$ state, a finding that has been corroborated by recent experimental results (Chen \emph{et al.}, 2021; Yao \emph{et al.}, 2022; Yao \emph{et al.}, 2023). An extension to a $C_2$ symmetric double-$\textbf{k}$ in the case of monoclinic Na$_3$Co$_2$SbO$_6$ has also been suggested (Gu \emph{et al.}, 2024). While there is currently still much debate in the community as to the true nature of these ground states, in light of these reports, we will review and provide detailed crystallographic descriptions of the possible multi-$\mathbf{k}$ structures that were proposed to emerge in the aforementioned systems. 

\begin{table}
\caption{List of honeycomb compounds with a zigzag AFM ground state}
\label{zizag_table}
\begin{tabular}{K{2.4cm}K{0.8cm}K{1.1cm}K{2.2cm}K{1.5cm}K{3.7cm}}\hline

Compound & $S_{\textrm{eff}}$ & Crystal SG & $\mathbf{k}$-vector & T$_{N}$ & Refs. \\ [0.5ex] \hline \hline
Na$_{2}$IrO$_{3}$ & $\nicefrac{1}{2}$ & \textit{C2/m} & (0, 1, $\nicefrac{1}{2}$) & 18~K & Ye \emph{et al.}, 2012 \\
Na$_{3}$Co$_{2}$SbO$_{6}$ & $\nicefrac{1}{2}$ & \textit{C 2/m} & ($\nicefrac{1}{2}$, $\nicefrac{1}{2}$, 0) & 6.7 - 8 K & Wong \emph{et al.}, 2016; Yan \emph{et al.}, 2019; Li \emph{et al.}, 2022 \\
Na$_{2}$Co$_{2}$TeO$_{6}$ & $\nicefrac{1}{2}$ & \textit{C 2/m} & ($\nicefrac{1}{2}$, $\nicefrac{1}{2}$,0)& 9.6~K & Dufault \emph{et al.}, 2023 \\
Li$_{3}$Ni$_{2}$SbO$_{6}$ & 1 & \textit{C 2/m} & ($\nicefrac{1}{2}$, $\nicefrac{1}{2}$, 0) & 15~K & Kurbakov \emph{et al.}, 2015; Zvereva \emph{et al.}, 2015 \\
Na$_{3}$Ni$_{2}$SbO$_{6}$ & 1 & \textit{C 2/m} & ($\nicefrac{1}{2}$, $\nicefrac{1}{2}$, 0) & 16~K & Zvereva \emph{et al.}, 2015; Werner \emph{et al.}, 2019 \\
Na$_{3}$Ni$_{2}$BiO$_{6}$ & 1 & \textit{C 2/m} & (0, 1, 0) & 10.4~K & Seibel \emph{et al.}, 2013; Shangguan \emph{et al.}, 2023 \\
Li$_{3}$Ni$_{2}$BiO$_{6}$ & 1 & \textit{C 2/m} & tbd & 5.5~K & Berthelot \emph{et al.}, 2012 \\
Cu$_{3}$Co$_{2}$SbO$_{6}$ & $\nicefrac{1}{2}$ & \textit{C 2/c} & (1, 0, 0) & 18.5~K & Roudebush \emph{et al.}, 2013 \\
Cu$_{3}$Ni$_{2}$SbO$_{6}$ & 1 &\textit{C 2/c} & (1, 0, 0) & 22.3 K & Roudebush \emph{et al.}, 2013 \\ 
Ag$_{3}$Co$_{2}$SbO$_{6}$ & $\nicefrac{1}{2}$ & \textit{C 2/m} & tbd & 21.2~K & Zvereva \emph{et al.}, 2016 \\
CoPS$_{3}$ & $\nicefrac{3}{2}$ & \textit{C 2/m} & (0, 1, 0) & 122~K & Wildes \emph{et al.}, 2017 \\
NiPS$_{3}$ & 1 & \textit{C 2/m} & (0, 1, 0)  & 155~K & Wildes, 2015 \\
NiPSe$_{3}$ & 1 & \textit{C 2/m} & tbd & 212~K & Basnet, 2022 \\ [0.5ex] \hline
Na$_{2}$Co$_{2}$TeO$_{6}$ & $\nicefrac{1}{2}$ & \textit{P6$_{3}$22} & ($\nicefrac{1}{2}$, 0, 0) & 27~K & Lefran\c{c}ois \emph{et al.}, 2016; Bera \emph{et al.}, 2017; Samarakoon \emph{et al.}, 2021 \\
K$_{2}$Co$_{2}$TeO$_{6}$ & $\nicefrac{1}{2}$ & \textit{P6$_{3}$mcm} & tbd & 12.3~K & Xu \emph{et al.}, 2023 \\
Na$_{2}$Ni$_{2}$TeO$_{6}$ & 1 & \textit{P6${}_{3}$mcm} & ($\nicefrac{1}{2}$, 0, $\nicefrac{1}{2}$) & 27 - 30~K & Samarakoon \emph{et al.}, 2021; Bera \emph{et al.}, 2022 \\
Na$_{2}$Ni$_{2}$TeO$_{6}$ & 1 & \textit{P6${}_{3}$mcm} & ($\nicefrac{1}{2}$, 0, 0) & 27.5~K & Karna \emph{et al.}, 2017; Kurbakov \emph{et al.}, 2020 \\
Na$_{2.4}$Ni$_{2}$TeO$_{6}$ & 1 & \textit{P6$_{3}$mcm} & ($\nicefrac{1}{2}$, 0, 0) & 25~K & Samarakoon \emph{et al.}, 2021 \\
K$_{2}$Ni$_{2}$TeO$_{6}$ & 1 & \textit{P6$_{3}$mcm} & tbd  & 23 - 27~K & Matsubara \emph{et al.}, 2020; Vasilchikova \emph{et al.}, 2022 \\ [0.5ex] \hline 
BaNi$_{2}$(AsO$_{4}$)$_{2}$ & 1 & \textit{R-3} & ($\nicefrac{1}{2}$, 0, $\nicefrac{1}{2}$) & 18~K & Regnault \emph{et al.}, 1980; Gao \emph{et al.}, 2021 \\
$\alpha$-RuCl$_{3}$ & $\nicefrac{1}{2}$ & \textit{R-3} & (0, $\nicefrac{1}{2}$, 1) & 6.3 - 14~K & Park \emph{et al.}, 2024; Cao \emph{et al.}, 2016 \\
Ru$_{0.9}$(Ir,Rh)$_{0.1}$Cl$_{3}$ & $\nicefrac{1}{2}$ & \textit{R-3} & (0, $\nicefrac{1}{2}$, 1) & 6~K & Morgan \emph{et al.}, 2024 \\
$\alpha$-RuBr$_{3}$ & $\nicefrac{1}{2}$ & \textit{ R-3} & (0, $\nicefrac{1}{2}$, 1) & 34~K &  Imai, 2022  \\
KNiAsO$_{4}$ & 1 & \textit{R-3} & ($\nicefrac{3}{2}$, 0, 0) & 19~K & Bramwell \emph{et al.}, 1994; Taddei \emph{et al.}, 2023 \\
KCoAsO$_{4}$ & $\nicefrac{1}{2}$ & \textit{R-3} & ($\nicefrac{3}{2}$, 0, 0) \par ($\nicefrac{1}{2}$, 0, $\nicefrac{1}{2}$) & 13~K & Pressley \emph{et al.}, 2024 \\ \hline
\end{tabular}
\end{table}

\section{Monoclinic honeycomb systems with zigzag-type magnetic structures.}

Among the earliest and most notable candidates for the Kitaev model are the A$_2$IrO$_3$ iridates (A = Na, Li, Cu), renowned for their intriguing magnetic and electronic behaviors. These phenomena result from the combined effects of crystal field interactions, spin-orbit coupling, and electron correlations. The A$_2$IrO$_3$ compounds crystallize in the monoclinic space group $C2/m$, and are characterized by a layered arrangement with edge-sharing  octahedra. Central to these layers are iridium (Ir$^{4+}$) ions with a pseudospin-1/2 ($J_{\textrm{eff}}$ = 1/2), situated at the core of the edge-sharing oxygen octahedra, which form a honeycomb pattern. The structure also includes three inequivalent cation (A$^{1+}$) sites. A notable feature of these materials is a certain level of disorder, often due to the mixing of cation sites, leading to heterogeneous magnetic ground states. In this class of materials, only Na$_2$IrO$_3$ was found to display a zigzag antiferromagnet order (Ye \emph{et al.}, 2012). The three structural polytypes of Li$_2$IrO$_3$ (known as $\alpha$-, $\beta$-  and $\gamma$- Li$_2$IrO$_3$) exhibit incommensurate magnetic orders with counter-rotating moments on nearest-neighbor sites (Williams, 2016; Biffin \emph{et al.}, 2014). Meanwhile, the magnetic ground state of Cu$_2$IrO$_3$ (Kenney \emph{et al.}, 2019) was shown to consist of coexistent static and dynamic (spin liquid) magnetic states originated from a mixed copper valence Cu$^+$/Cu$^{2+}$. For H$_3$LiIr$_2$O$_6$ (Kitagawa, 2018) and Ag$_3$LiIr$_2$O$_6$ (Chakraborty \emph{et al.}, 2021), experimental evidence suggest the existence of a quantum spin liquid state and coexisting incommensurate N\'{e}el and stripe ordered magnetic domains, respectively.

\begin{figure}
\includegraphics[width=5in]{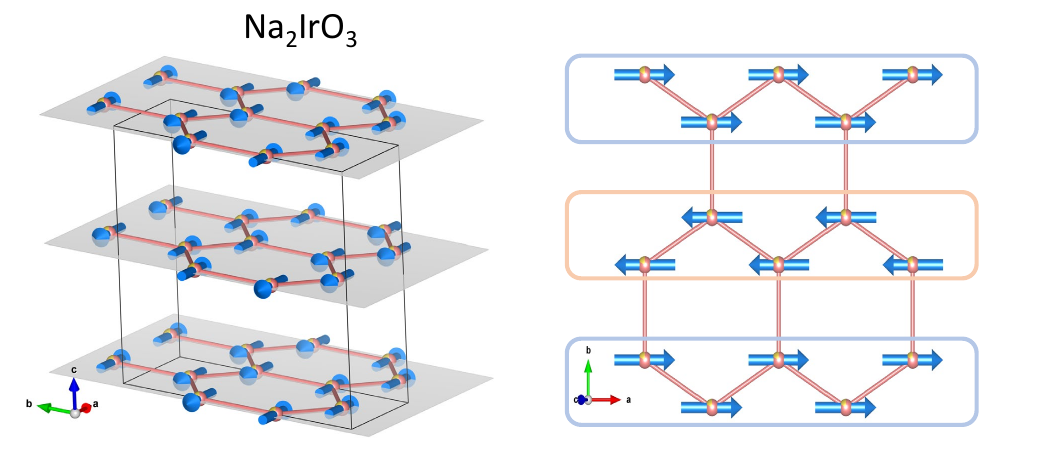}
\caption{ Magnetic structure of Na$_{2}$IrO$_{3}$ and view along $c$* of the honeycomb layer consisting of alternating ferromagnetic zigzag chains. Magnetic structures were drawn using VESTA software (Momma \& Izumi, 2011)}
\label{fig:nairo}
\end{figure}

The zigzag magnetic structure of Na$_2$IrO$_3$ was unveiled by Ye \textit{et al.} (2012) from single crystal neutron diffraction data. The magnetic propagation vector was identified as $\mathbf{k}$=(0, 1, $\nicefrac{1}{2}$), indicating the possibility of either stripy or zigzag-type order, below the N\'{e}el temperature T$_N$ = 18.1(2) K. The zigzag arrangement with ferromagnetic chains running along the $a$-axis, depicted in Fig.~\ref{fig:nairo}, was inferred from the relative intensities of three magnetic reflections. A comprehensive crystallographic description of the magnetic structure of Na$_2$IrO$_3$ is provided in Table~\ref{table_mono1} in the Appendix. This magnetic structure is described by the magnetic space group $C_c  2/m$ (\# 12.63) in a magnetic cell (a ,b, 2c). The magnetic space group confines the magnetic moment to be in the $ac$-plane. However, data from magnetic susceptibility, polarized x-ray measurements, and neutron experiments suggest that the moment is primarily oriented along the $a$-axis. The fitted neutron scattering intensities resulted in a magnetic moment of 0.22(1)$\mu_B$/Ir, which is significantly smaller than expected for the expected $J_{\textrm{eff}}$ = 1/2 state. This reduced moment is attributed to the strong hybridization between the Ir 5$d$ orbitals and the ligand oxygen 2$p$ orbitals.

Another important category of compounds featuring honeycomb magnetic lattices is characterized by the general formula $A_{2,3}M_{2}M'O_{6}$, where $A$ represents an alkali metal (Na, Li, K), $M$ denotes a metal transition ion such as Co$^{2+}$ or Ni$^{2+}$ and $M'$ is a heavy metal such as Sb, Te, or Bi. Kanyolo \textit{et al.} (2023) offered an exhaustive review of honeycomb layered oxides within this group, encompassing detailed synthesis methods. Viciu \emph{et al.} (2007) identified in 2007 the layered compounds Na$_{3}$Co$_{2}$SbO$_{6}$ and Na$_{2}$Co$_{2}$TeO$_{6}$, which are akin to the layered Na$_x$CoO$_2$, albeit with a third of the Co sites substituted by the non-magnetic ions Sb$^{5+}$/Te$^{6+}$. These compounds exhibit complete ordering of Co and Sb$^{5+}$/Te$^{6+}$ ions, forming a honeycomb lattice. However, the Na distribution amidst the honeycomb layers is markedly disordered. Na$_{3}$Co$_{2}$SbO$_{6}$ was found to have a single-layer monoclinic structure (space group $C2/m$), whereas Na$_{2}$Co$_{2}$TeO$_{6}$ possesses a two-layer hexagonal structure ($P6_322$). Notably, a monoclinic ($C2/m$) polymorph of Na$_{2}$Co$_{2}$TeO$_{6}$ was recently documented by Dufault \textit{et al.} (2023).

The zigzag-type magnetic structure of Na$_{3}$Co$_{2}$SbO$_{6}$ has been corroborated by multiple research groups, as defined by a propagation vector $\mathbf{k}$ =($\nicefrac{1}{2}$, $\nicefrac{1}{2}$, 0) and a magnetic space group $P_S \bar{1}$ (\# 2.7). The magnetic symmetry permits the orientation of the magnetic moment along any crystallographic axis. Initial powder neutron diffraction study suggested that the Co magnetic moments align parallel to the $c$-axis (Wong \emph{et al.}, 2016). However, this model was subsequently contested by a single-crystal study (Yan \emph{et al.}, 2019), which proposed a zigzag configuration with moments of 0.9 $\mu_{B}$ directed along the $b$-axis. Confinement of the moment within the $ab$ plane, orthogonal to the wave vector, was later substantiated by further single-crystal studies employing neutron polarization analysis (Li \emph{et al.}, 2022; Gu \emph{et al.}, 2024). Additionally, in-plane magnetic field experiments suggested a potential non-collinear double-$\mathbf{k}$ structure, indicative of XXZ easy-plane anisotropy, as opposed to Kitaev anisotropy (Gu \emph{et al.}, 2024). 

The magnetic structure of the monoclinic variant of Na$_{2}$Co$_{2}$TeO$_{6}$ was refined using powder neutron diffraction data, with magnetic intensities indexed by the identical $\mathbf{k}$ vector ($\nicefrac{1}{2}$, $\nicefrac{1}{2}$, 0) (Dufault \emph{et al.}, 2023). The Rietveld refinement yielded an out-of-plane canted moment with the components $m_x$=0.48, $m_y$=1.5(1) and $m_z$=1.2(1)$\mu_B$. The single-$\mathbf{k}$ zigzag-like magnetic structure models of both Sb and Te cobaltates are shown in Fig.~\ref{fig:mono1}. Detailed crystallographic information is given in Table~\ref{table_mono1} of the Appendix.

The past decade has seen a rapid increase in the number of new compounds within the $A_{2,3}M_{2}M'O_{6}$ class. The list of $C2/m$ monoclinic, single-layer, honeycomb systems with verified zigzag magnetic structures now include several Ni-based ($S$ = 1) compounds such as Li$_{3}$Ni$_{2}$SbO$_{6}$, Na$_{3}$Ni$_{2}$SbO$_{6}$ and Na$_{3}$Ni$_{2}$BiO$_{6}$ (Fig.~\ref{fig:mono1}). The Ni-antimonides compounds are characterized by a magnetic supercell (2a, 2b, c), which originates from the propagation vector ($\nicefrac{1}{2}$, $\nicefrac{1}{2}$, 0). The ordered magnetic moments are slightly inclined from the $c$-axis by about 17$\pm$1$^{\circ}$ towards the $a$-axis, rendering the moments nearly orthogonal to the honeycomb plane. The moment magnitudes are in the range 1.6 - 1.8 $\mu_B$, marginally less than the expected value for a $S$ = 1 system. The Bi-analog Na$_{3}$Ni$_{2}$BiO$_{6}$ exhibits ordered magnetic moments also perpendicular to the honeycomb plane, yet the zigzag chain orientation differs (Seibel \emph{et al.}, 2013). Interestingly, the propagation vector $\mathbf{k}$ = (0, 1, 0), defining the magnetic order in this material, results in ferromagnetic chains along $a$-axis and alternating directions along $b$-axis. This model is described using the magnetic space group $P_C 2_1/m$ (\# 11.57). Crystallographic description of magnetic structure is provided in Table~\ref{table_mono2}. Given the small distortion of the honeycomb lattice in the monoclinic structure, it is inferred that the FM zigzag chains along $a$-axis consist of uniform Ni-Ni bonds, whereas the Sb and Te compounds feature chains with alternating Ni-Ni bonds. The N\'{e}el temperature typically declines when transitioning from Sb to Bi, with the same alkali. A recent study revealed that when subjected to an external magnetic field, Na$_{3}$Ni$_{2}$BiO$_{6}$ exhibits a one-third magnetization plateau. This is indicative of a ferrimagnetic state, which implies the presence of bond-anisotropic Kitaev interactions in this phase (Shangguan \emph{et al.}, 2023). This discovery has attracted significant interest, broadening the scope of fractional magnetization plateau phase research to honeycomb-lattice compounds with $S$ = 1. Meanwhile, the sister compound Li$_{3}$Ni$_{2}$BiO$_{6}$ (Berthelot \emph{et al.}, 2012) has yet to be thoroughly investigated and awaits neutron scattering studies to confirm zigzag magnetic ordering.

\begin{figure}
\includegraphics[width=5in]{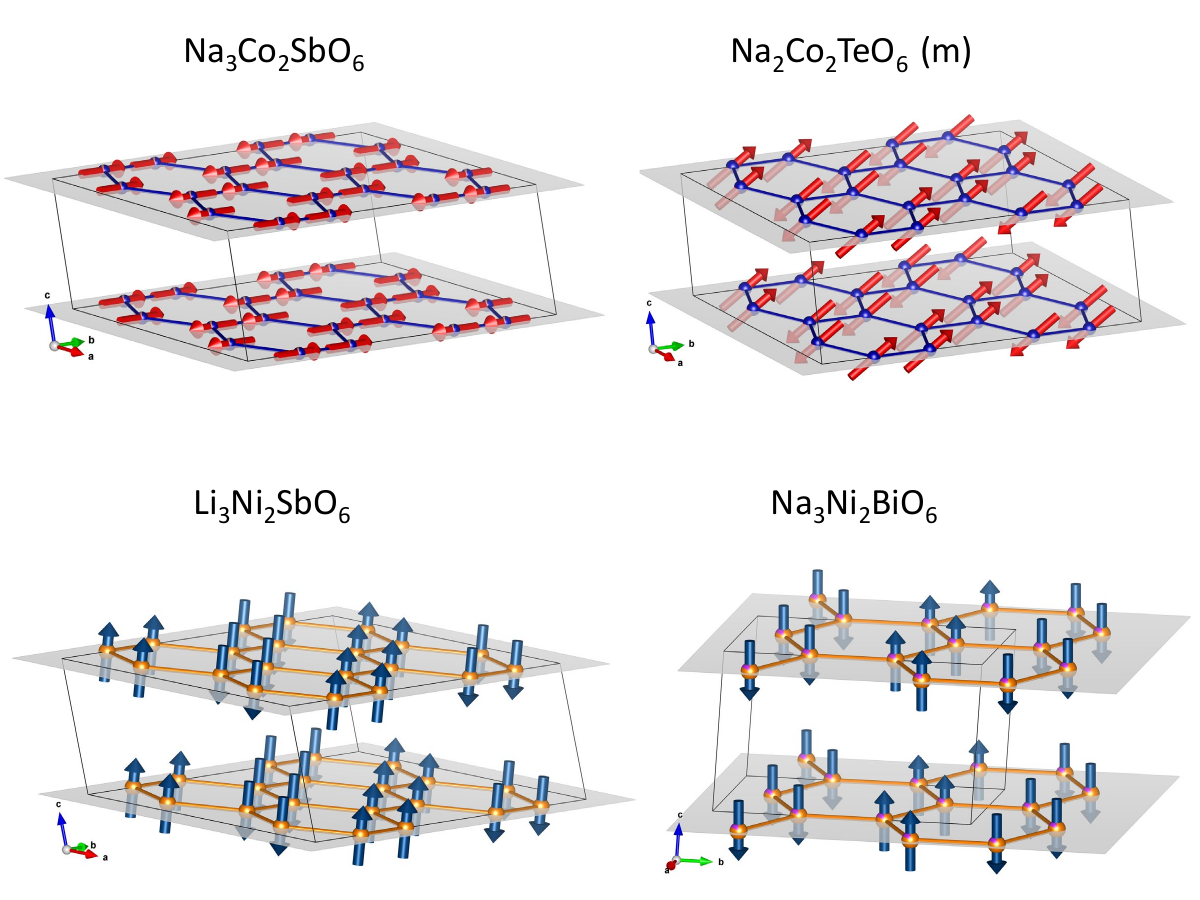}
\caption{Zigzag magnetic structure of monoclinic honeycomb systems $A_{2,3}$M$_{2}$M'O$_{6}$ with $A$ = Na, Li and M =  Co and Ni. The Co moments are mainly confined to the $ab$ plane, while the Ni moments are orthogonal to the honeycomb plane. The magnetic structures of Sb and Te compounds are defined by $\mathbf{k}$ = ($\nicefrac{1}{2}$, $\nicefrac{1}{2}$, 0), forming ferromagnetic zigzag chains running along the diagonal direction. In contrast, the magnetic structure of Na$_{3}$Ni$_{2}$BiO$_{6}$ is defined by a propagation vector $\mathbf{k}$ = (0, 1, 0), which produce ferromagnetic chains along $a$-axis.}
\label{fig:mono1} 
\end{figure}

\begin{figure}
\includegraphics[width=5in]{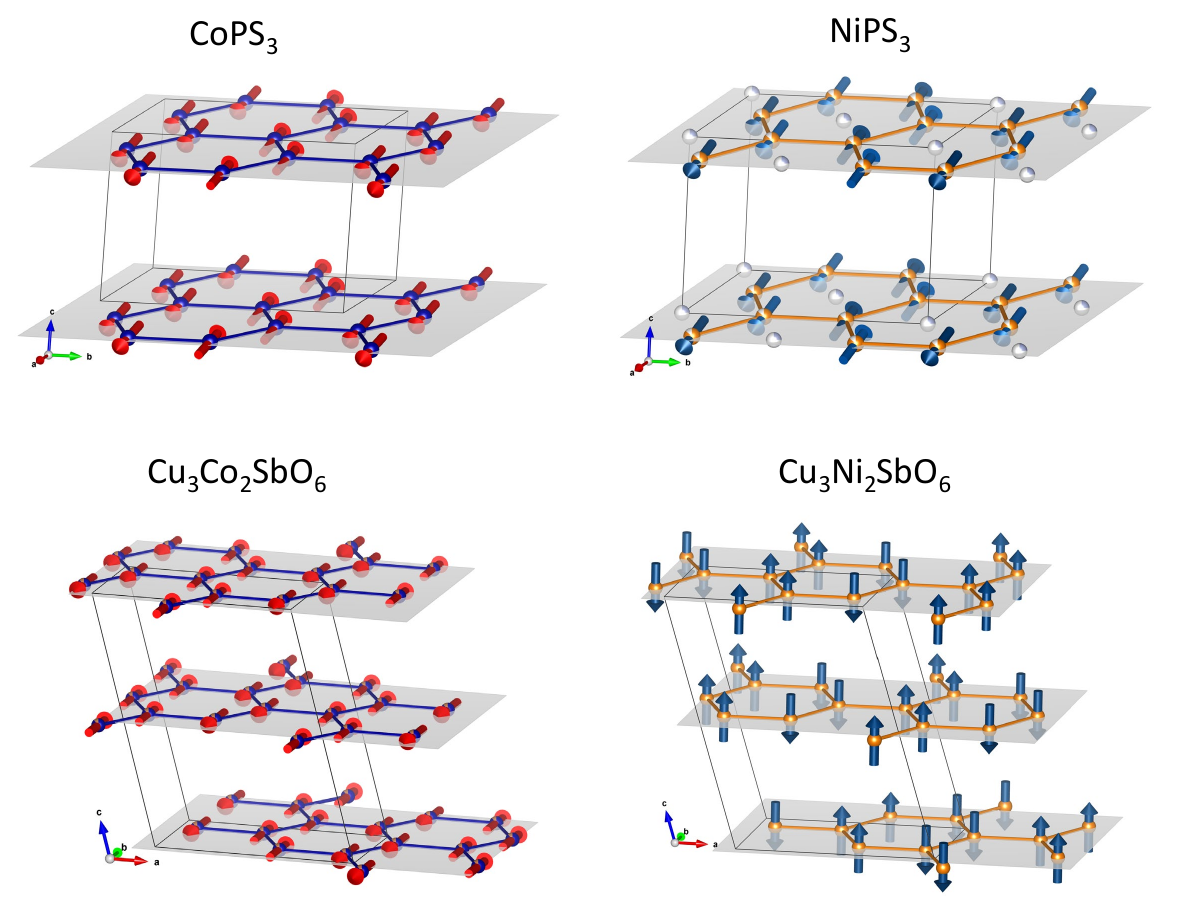}
\caption{Zigzag magnetic structure of $M$PS$_{3}$ and delafossite-derived Cu$_{3}M_{2}$SbO$_{6}$, where $M$ =  Co and Ni.}
\label{fig:mono2} 
\end{figure}

The delafossite-type structures have a similar formula, $A_{3}M_{2}M'O_{6}$, where the monovalent ion ($A$) can be either Cu$^{1+}$ or Ag$^{1+}$. Unlike alkali-metal systems, these structures feature linear O-Cu-O coordination instead of octahedral. Notable examples include the Cu$_{3}$Co$_{2}$SbO$_{6}$ and Cu$_{3}$Ni$_{2}$SbO$_{6}$ compounds, which exhibit a monoclinic distortion of the hexagonal delafossite polytype. Their structure is described by the space group $C2/c$ and contain two honeycomb layers per unit cell. Antiferromagnetic ordering occurs at 22.3 K for Ni and 18.5 K for Co variants. The magnetic structures are defined by a propagation vector $\mathbf{k}$ = (1, 0, 0) and involve FM zigzag chains along the $b$-axis, antiferromagnetically coupled to neighboring chains (see Fig.~\ref{fig:mono2}). In the Co delafossite, magnetic moments align parallel to the $b$-axis, while in the Ni compound, they are nearly orthogonal to the honeycomb plane. Adjacent honeycomb layers exhibit antiferromagnetic coupling. These magnetic models were described within same monoclinic crystal frame by the magnetic space group $P_C 2_1/c$ (for Co) and $P_C 2/c$ (for Ni). Additional structural details are summarized in Table~\ref{table_mono3}. Notably, the Ni compound exhibits a second polytype described by the hexagonal space group $P6_3$, with its magnetic order defined by a propagation vector $\mathbf{k}$ = ($\nicefrac{1}{2}$, 0, 0). Another identified delafossite-honeycomb material is Ag$_{3}$Co$_{2}$SbO$_{6}$ (Zvereva, 2016). Although ab-initio GGA+U calculations predict a zigzag-type magnetic structure, no experimental confirmation exists to date.

The metal phosphorous trichalcogenides ($M$P$X_3$) formed from transition metals (M) and chalcogen atoms ($X$ = S, Se, Te) hold promise as single-layer van der Waals materials with potential applications in spintronics. The crystal structure has monoclinic symmetry with space group $C2/m$. Specifically, CoPS$_3$ and NiPS$_3$ were found to exhibit zigzag-type structures, stemming from dominant antiferromagnetic third-neighbor exchange coupling $J_3$ (Wildes \emph{et al.}, 2015; 2017; Kim \emph{et al.}, 2020; Wildes \emph{et al.}, 2022; 2023; Scheie \emph{et al.}, 2023). Both compounds order with a propagation vector $\mathbf{k}$ =(0, 1, 0), below the N\'{e}el temperature of  122 K (for Co) and 155 K (for Ni). The magnetic structures are depicted in Fig.~\ref{fig:mono2}. The magnetic structure of CoPS$_3$ was refined from neutron single crystal data using a model described by the magnetic space group $P_C 2_1/m$ (\# 11.57).  Refinement results allowed the moments to have components along the $a$ and $c$ axes, as constrained by symmetry. The total moment was found to be 3.3 $\mu_B$, with the components $m_x$=3.13(8) and $m_z$=-0.6(1) $\mu_B$, consistent with an absence of an spin orbit exciton in neutron scattering suggesting a pure $S$ = 3/2 state (Kim \emph{et al.}, 2020). Structural studies on NiPS$_3$ reported a site disorder between a main Ni1 ($4g$) and the minority Ni2 ($2a$) sites, with the ratio of approximately 88:12. The magnetic structure was refined from neutron single crystal data. The zigzag-type structure involves three irreducible representations (Irreps) to accommodate the magnetic ordering of both sites. The model, defined by the magnetic space group $P_S 1$ (\# 1.3), breaks all the point-group symmetry operations. While the magnetic refinement of the Ni1 ($4g$) site appeared to be robust, there has been less confidence in the moment direction for the minority $2a$ site. The moment magnitudes were found to be 1.05 $\mu_B$ ($m_x$=0.94(3), $m_y$=0, $m_z$=-0.26(4)) for Ni1, and 1.06 $\mu_B$ ($m_x$=1.0(5), $m_y$=0.5(6), $m_z$=0.0(9)) for Ni2 sites. The moment direction predominantly aligns with the $a$-axis, consistent with spin wave analyses that revealed a significant XY-like anisotropy accompanied by a minor uniaxial component (Wildes \emph{et al.}, 2022; Scheie \emph{et al.}, 2023). One shall note that considering only the Ni1 order, the zigzag structure can be equally well described by the magnetic space group $P_C 2_1/m$, similar to that proposed for the Co compound. For further details, refer to Table~\ref{table_mono3}.

\section{Hexagonal honeycomb systems with zigzag-type magnetic structures.}

\begin{figure}
\includegraphics[width=5in]{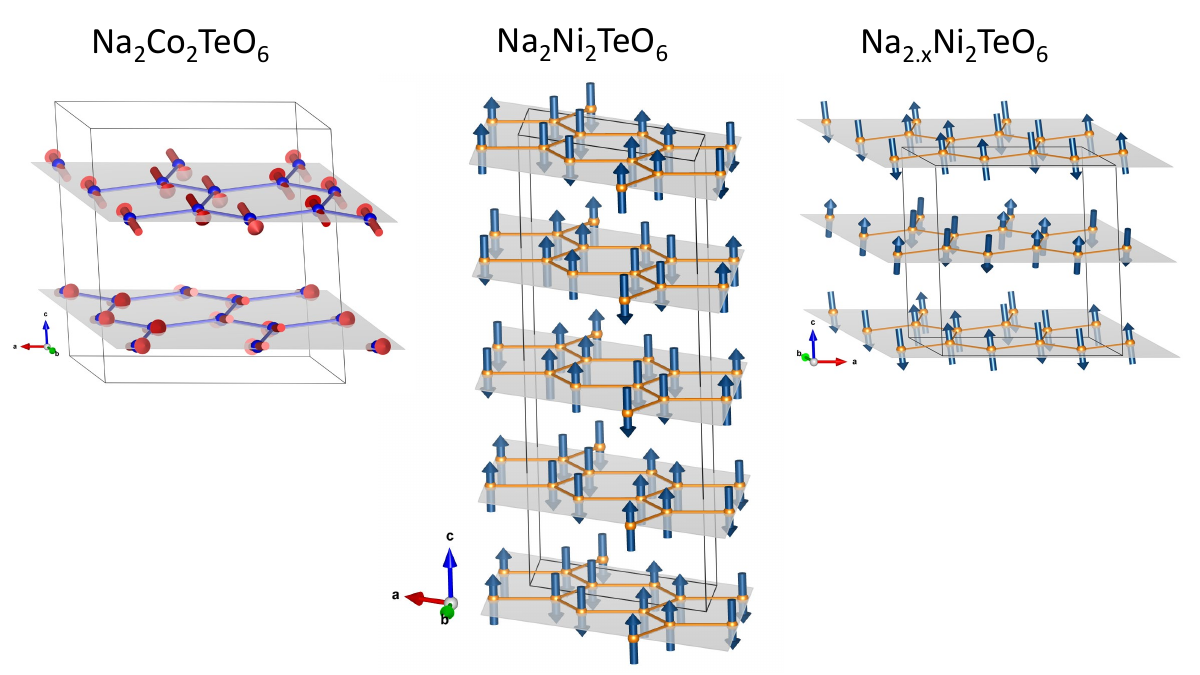}
\caption{Zigzag magnetic structure of hexagonal Na$_{2}$Co$_{2}$TeO$_{6}$ and two layer stacking variants of Na$_{2}$Ni$_{2}$TeO$_{6}$ defined by $\mathbf{k}$ = ($\nicefrac{1}{2}$,~0,~$\nicefrac{1}{2}$) and ($\nicefrac{1}{2}$,~0,~0) wave-vectors}
\label{fig:hex}
\end{figure}

The hexagonal honeycomb systems known to exhibit zigzag magnetic ordering belong to the $A_{2}M_{2}M'O_{6}$ class with $M'$ = Te$^{6+}$. In particular, Na$_{2}$Co$_{2}$TeO$_{6}$ presents an exciting opportunity for exploring quantum magnetism and exotic magnetic states. Its zigzag magnetic order and potential connection to the Kitaev model have made it a captivating subject of study within the condensed matter community. Its crystal structure is described by $P6_322$ space-group and the successive Co honeycomb planes are shifted by a [$\nicefrac{1}{3}$, $\nicefrac{2}{3}$, 0] translation. The honeycomb networks are formed by two nonequivalent Co sites, Co1 in the $2b$ (0, 0, $\nicefrac{1}{4}$) site and Co2 in the $2d$ ($\nicefrac{2}{3}$, $\nicefrac{1}{3}$, $\nicefrac{1}{4}$). Crystal electric field excitations provided evidence that Co$^{2+}$ ions have a spin-orbital entangled $S_{\textrm{eff}}$ = 1/2 state (Songvilay \emph{et al.}, 2020). Magnetic susceptibility showed a long range magnetic order at about 27~K followed by with two spin reorientation transitions at 15 and 5~K. (Lefran\c{c}ois \emph{et al.}, 2016; Bera \emph{et al.}, 2017; Samarakoon \emph{et al.}, 2021) The magnetic order is described by the propagation vector $\mathbf{k}$ = ($\nicefrac{1}{2}$,~0,~0). The zigzag magnetic model, described by the magnetic space group $P_C2_12_12_1$  (\# 19.29) in a (2a,b,c) magnetic cell base, provided the best description of neutron scattering intensities. Ferromagnetic chains align along the $b$-direction (orthogonal to $\mathbf{k}$), and the magnetic moments lie in the $bc$-plane. The magnetic moments alternate their directions in successive layers. The out-of-plane component ($m_z$) is particularly relevant in the context of the Kitaev-type anisotropic bond-directional couplings model. Refined magnetic moments for the two Co positions were $m_y$ = 2.07(7)$~\mu_B$ and $m_z$ = 0.5(2)$~\mu_B$ for Co1, and $m_y$ =1.95(10)$~\mu_B$ and $m_z$ = 0.5(2)$~\mu_B$ for Co2 (Samarakoon \emph{et al.}, 2021). Other reported values, when only in-plane components were considered, were: $m_{Co1}$ = 2.7$~\mu_B$, and $m_{Co2}$=2.45$~\mu_B$ (Bera \emph{et al.}, 2017). Fig.~\ref{fig:hex} illustrates the magnetic structure, and Table~\ref{table_hex} provides a tabulated description. A 3-$\mathbf{k}$ magnetic structure model was recently proposed by several studies (Chen \emph{et al.}, 2021; Yao \emph{et al.}, 2022, 2023; Kr\"{u}ger \emph{et al.}, 2023), and further details will be discussed in a subsequent section. The magnetic field evolution of the magnetic structure of Na$_{2}$Co$_{2}$TeO$_{6}$ was recently explored by Yao \emph{et al.}, 2023 and Bera \emph{et al.}, 2023. 

A single crystal of K$_{2}$Co$_{2}$TeO$_{6}$ compound was recently synthesized and its structural analysis revealed a $P6_3/mcm$ space group (Xu \emph{et al.}, 2023). Unlike the Na variant, the honeycomb layers in K$_{2}$Co$_{2}$TeO$_{6}$ have a different stacking arrangement, with Co atoms directly stacked on top of each other. The honeycomb lattices consist of a single Co site ($4d$), and K ions are distributed disorderly between the honeycomb layers. The system exhibits antiferromagnetic order below 12.3~K, although no magnetic diffraction studies have been reported to date.

The Ni-based compounds  A$_{2}$Ni$_{2}$TeO$_{6}$ are isostructural to K$_{2}$Co$_{2}$TeO$_{6}$. Notably, the magnetic order of these systems is influenced by alkali-ion sites occupancies. Samarakoon \textit{et al.} (2021) reported a propagation vector $\mathbf{k}$ = ($\nicefrac{1}{2}$,~0,~$\nicefrac{1}{2}$) for the stoichiometric Na$_{2}$Ni$_{2}$TeO$_{6}$ and $\mathbf{k}$ = ($\nicefrac{1}{2}$,~0,~0) for the doped Na$_{2.4}$Ni$_{2}$TeO$_{6}$. Earlier neutron diffraction studies on Na$_{2}$Ni$_{2}$TeO$_{6}$ reported a ($\nicefrac{1}{2}$,~0,~0) wave vector (Karna \emph{et al.}, 2017; Kurbakov \emph{et al.}, 2020), while Bera \textit{et al.} observed a mixture of the two $\mathbf{k}$ vectors. This discrepancy likely arises from variations in Na composition and sample homogeneity. The key difference in the ordered state lies in the stacking sequence of adjacent honeycomb layers that changes from a \textit{A-B-A-B} type to \textit{A-A-B-B}, where \textit{A} and \textit{B} represent opposite moment directions. This suggests that the out-of-plane coupling is highly sensitive to Na amount, and distribution, within the Na monolayer.

For the undoped material, the magnetic structure is described using the magnetic space group $I_amm2$ (\# 44.234) in the supercell (2a, b, 2c). Moments align parallel to the $c$-axis, forming a zigzag structure with ferromagnetic chains along the $b$ direction. The refined static moment is 1.55(6)~$\mu_B$, with an in-plane component less than 0.05~$\mu_B$ (Samarakoon \emph{et al.}, 2021). In contrast, the zigzag-type magnetic structure of Na$_{2.4}$Ni$_{2}$TeO$_{6}$ is described by the magnetic space group $P_A nma$ (\# 62.453) on the magnetic lattice (2a, b, c). The ordered moment exhibits canting away from the $c$-axis, with the magnetic components $m_y$=0.7(1)$~\mu_B$ and $m_z$=1.30(5)$~\mu_B$. Graphical representations of Na$_{2.x}$Ni$_{2}$TeO$_{6}$ magnetic structures are shown in Fig.~\ref{fig:hex}. The magnetic structure of the potassium variant K$_{2}$Ni$_{2}$TeO$_{6}$, has not yet been examined through neutron scattering. However, spin-polarized DFT + U calculations indicate that the honeycomb layers of this compound also adopts a zigzag order (Vasilchikova \emph{et al.}, 2022).

\section{Honeycomb materials with rhombohedral lattice}

The honeycomb $\alpha$-RuCl${_3}$ has remained arguably one of the most topical magnets of the last decade. After numerous experimental and theoretical investigations $\alpha$-RuCl${_3}$ has emerged as a leading candidate material for observing Kitaev physics. The observation of a suppression of magnetic order under an applied magnetic field within the honeycomb plane above 7.3 T (Banerjee \emph{et al.}, 2018), has led to the proposal that $\alpha$-RuCl${_3}$ is a proximate Kitaev QSL in a narrow field range of 7.3 - 10 T. Fractionalized excitations have been reported in neutron scattering experiments (Banerjee \emph{et al.}, 2016,2017,2018; Do \emph{et al.}, 2017; Balz, 2019), Raman scattering (Dandilands \emph{et al.}, 2015), and THz spectroscopy (Little \emph{et al.}, 2017), along with reports of quantized thermal Hall plateaus matching the half-integer value expected from Kitaev's model (Kasahara \emph{et al.}, 2018; Yokoi \emph{et al.}, 2021). Despite strong support for Kitaev physics, there have been several conflicting reports from various groups on $\alpha$-RuCl${_3}$ as a candidate material, and to this day discussion of the nature of the field induced disordered phase remains a centerpiece in the field. As work on $\alpha$-RuCl${_3}$ can easily spawn its own rich review article, in this work we focus mostly on its low temperature crystalline and magnetic structure.

\begin{figure}
\includegraphics[width=5in]{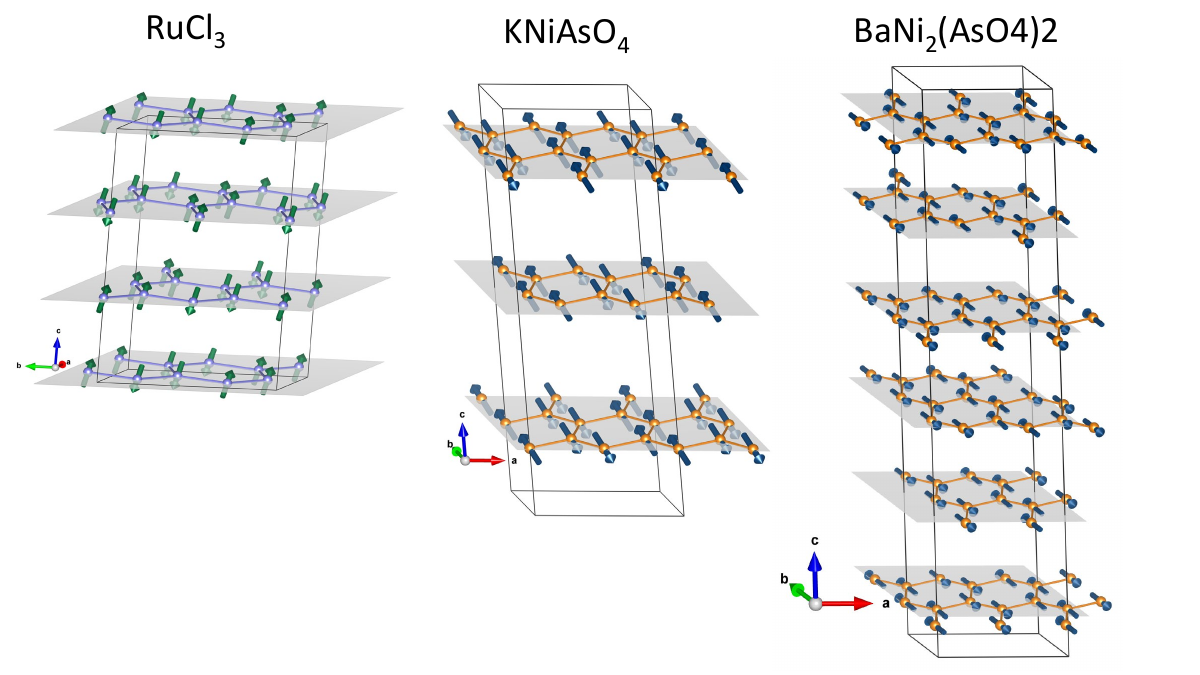}
\caption{Zigzag magnetic structure observed in honeycomb material $\alpha$-RuCl${_3}$, KNiAsO$_4$ and BaNi$_{2}$(AsO$_{4}$)$_2$ with trigonal $R \bar{3}$ parent symmetry.}
\label{fig:rho}
\end{figure}

The low-temperature crystal structure of $\alpha$-RuCl${_3}$ has been a topic of extensive debate due to its structural fragility and apparent sensitivity to the crystal growth method (Johnson \emph{et al.}, 2015; Sears \emph{et al.}, 2015; Cao \emph{et al.}, 2016). The antiferromagnetic transition temperature, falls within the range of 6–14 K and is also highly sensitive to the specific details of the sample synthesis and handling, with high-quality stacking fault-free samples showing a single T$_N$ = 7 K. Though all studies of the room temperature structure confirm a monoclinic $C 2/m$ structure, multiple groups report a strongly first-order transition around 150 K (Kubota \emph{et al.}, 2015; Lampen-Kelley \emph{et al.}, 2018, Gass \emph{et al.}, 2020), below which $\alpha$-RuCl${_3}$ adopts a BiI$_3$-type structure with an ideal honeycomb lattice described by the $R \bar{3}$ space group (Mu \emph{et al.}, 2022; Park \emph{et al.}, 2024). The magnetic structure exhibits an alternating FM zigzag pattern, but uncertainties related to its low-temperature crystal symmetry and the prevalence of structural domains have hindered a precise determination of the magnetic moment orientation. Park \emph{et al.} (2024) reported a magnetic ordering with $\mathbf{k}$ = (0, $\nicefrac{1}{2}$,~1) and a refined magnetic moment of 0.73(3) $\mu_B$ per Ru atom. The magnetic moments lie within the $ac$-plane, tilted at an angle of 48(3)$^{\circ}$ from the hexagonal plane (see Fig.~\ref{fig:rho}). This structure is described with the magnetic space group $P_S \bar{1}$ for a cell doubled along the $b$-axis (a, 2b, c).

Chemical substitution with non-magnetic ions is an attractive strategy for tuning the magnetic behaviour of $\alpha$-RuCl${_3}$ and, potentially, for inducing a quantum spin-liquid state. In this context, Ru$_{0.9}R_{0.1}$Cl$_3$ ($R$ = Rh$^{3+}$ and Ir$^{3+}$) single crystals were prepared and characterized using neutron diffraction (Morgan \emph{et al.}, 2024). The study revealed that the structural transition from the monoclinic $C2/m$ to the trigonal $R\bar{3}$ crystal structure persists, though at a lower temperature of approximately 70~K. The N\'{e}el temperature is also slightly suppressed to around 6 K in the Ir-substituted compound, which exhibits a much reduced ordered moment of 0.32(5) $\mu_B$/Ru and a smaller canting angle of 15(4)$^{\circ}$.

A similar magnetic structure was also reported for the RuBr$_3$, which crystallizes in the $R \bar{3}$ structure and shows no structural transition. At 34 K, it undergoes a transition to the zigzag magnetic order characterized by the propagation vector $\mathbf{k}$ = (0, $\nicefrac{1}{2}$,~1) (Imai \emph{et al.}, 2022). The magnitude of the magnetic moment is determined to be 0.74(12)$\mu_B$ and its angle from the $ab$-plane is 64(12)$^{\circ}$. The crystallographic details of $\alpha$-RuCl${_3}$ and RuBr$_3$ are provided in Table~\ref{table_rho}.

Another Kitaev physics candidate is the spin-1 honeycomb compound KNiAsO$_4$, which orders magnetically below 19 K. It crystallizes in the trigonal $R\bar{3}$ space group, featuring edge-sharing NiO$_6$ octahedra that form layered structures. Each NiO$_6$-octahedral layer is enclosed by two AsO$_4$-tetrahedral sheets, and the interlayers contain offset triangular lattice K sheets. Early neutron diffraction work suggested an ordering vector of $\mathbf{k}$ = ($\nicefrac{1}{2}$,~0,~0) with an AFM chain structure, that results in a stripy-type structure (Bramwell \emph{et al.}, 1994). New, higher-resolution neutron powder measurements revealed a different propagation vector $\mathbf{k}$ = ($\nicefrac{3}{2}$,~0,~0) due to the observed extinction rule -h + k + l = 3n. The determined magnetic structure features FM zigzag chains along $a$-axis, described using the magnetic space group $P_S\bar{1}$ with doubled $a$ lattice parameter. The magnetic symmetry allows the magnetic moments to have non-zero components in all three crystallographic directions, and the refined components were $m_x$=-0.2(1), $m_y$ = 1.6(1) and $m_z$ = 0.7(1)$~\mu_B$. The magnetic moments mainly lie in the $ab$ plane with a small component along $c$ to produce a 22$^{\circ}$ canting, as depicted in Fig.~\ref{fig:rho}. The modeling of spin wave spectra indicates that the extended Kitaev spin Hamiltonian is essential for reproducing the observed canted zigzag-type order (Taddei \emph{et al.}, 2023). In the more recent study, the novel compound  KCoAsO$_4$ was added to this series. The static and dynamic magnetic properties were characterized, revealing a disordered ground state with two types of stacking sequences for the honeycomb layers (Pressley \emph{et al.}, 2024). Two sets of magnetic reflections indexed by ($\nicefrac{3}{2}$,~0,~0) and ($\nicefrac{1}{2}$,~0,~$\nicefrac{1}{2}$)~ $\mathbf{k}$-vectors were modelled using a zigzag-type magnetic arrangement. Interestingly, the Co magnetic moment was determined to be nearly parallel to the $c$-axis, with a small canting of about 7$^{\circ}$. This suggests opposite magnetic anisotropies for the Ni and Co compounds, but in a reversed manner compared to what was observed in the $A_{2,3}M_{2}M'O_{6}$ series.

A third family of honeycomb material with rhombohedral structure is Ba$M_2$($X$O$_4$)$_2$ ($M$ = Ni, Co, and X = P, As, V). These layered compounds consist of staggered honeycomb layers formed by edge-sharing $M$O$_6$ octahedra, separated by non-magnetic Ba and $X$O$_4$ layers. The magnetic properties and spin-Hamiltonians were investigated by Regnault and colleagues approximately four decades ago (Regnault \emph{et al.}, 1977, 1980). Each material exhibits a different magnetic propagation vectors: $\mathbf{k}$ = ($\nicefrac{1}{2}$,~0,~$\nicefrac{1}{2}$) for BaNi$_2$(AsO$_4$)$_2$, (0,~0,~$\nicefrac{3}{2}$) for BaNi$_2$(VO$_4$)$_2$ and (0, 0, 0) for BaNi$_2$(PO$_4$)$_2$. Among the compounds in this class, Ni-arsenate is the only one that exhibits a zigzag magnetic structure. On the other hand, the vanadate and phosphate variants display N\'{e}el-type order, where the magnetic moments couple antiferromagnetically to their three nearest neighbors. The zigzag magnetic order in BaNi$_2$(AsO$_4$)$_2$ follows the symmetry of the magnetic space group $P_S \bar{1}$ (\# 2.7) inside the magnetic lattice (2a, b, 2c). The magnetic moment was determined to be 2.0(2)~$\mu_B$ along $b$-axis, although the symmetry permits canting away from that axis. A much more complicated magnetic structure was observed in BaCo$_2$(AsO$_4$)$_2$ (Regnault \emph{et al.}, 1977). This system orders with an incommensurate propagation vector $\mathbf{k}$ = (0.261, 0, -1.31) and the inferred helical order consists of stacking weakly coupled quasi-ferromagnetic chains (Regnault \emph{et al.}, 2018). Weak in-plane magnetic fields can suppress the long-range magnetic order, suggesting it as a potential quantum spin liquid candidate (Halloran \emph{et al.}, 2023).

\section{Multiple-$\mathbf{k}$ magnetic structures derived from zigzag structures}

While the majority of magnetic structures involve a single propagation vector, there are cases when several $\mathbf{k}$ vectors can operate concomitantly to form fascinating multi-$\mathbf{k}$ structures. Theoretical investigations indicated that multi-$\mathbf{k}$ states can become energetically favorable over a single-$\mathbf{k}$ zigzag state when symmetry-allowed nonbilinear exchange perturbations are present (Kr\"{u}ger, 2023, Wang, 2023). Distinguishing between single-$\mathbf{k}$ and multi-$\mathbf{k}$ ordered states presents a significant experimental challenge. In crystals, multiple domains may be ordered according to different single $\mathbf{k}$ domains, resulting in the same contribution to magnetic scattering as the multi-$\mathbf{k}$ structure. To resolve ambiguity, scattering experiments on high-quality single crystals under external uniaxial stress or magnetic fields are necessary. These conditions can favor the population of a specific $\mathbf{k}$-domain to clarify the structure. A 1-$\mathbf{k}$ zigzag structure is expected to have one type of domain noticeably depopulated after a magnetic-field training, but a multi-$\mathbf{k}$ state will maintain the original ratio of the magnetic Bragg peaks. A recent neutron diffraction study on a twin-free crystal of monoclinic Na$_{3}$Co$_{2}$SbO$_{6}$, subjected to strong in-plane magnetic fields along a low-symmetry direction, revealed the presence of double-$\mathbf{k}$ order (Gu \emph{et al.}, 2024). On the other hand, when magnetic fields were applied along the $a$-axis of a hexagonal Na$_{2}$Co$_{2}$TeO$_{6}$ single crystal, evidence emerged of a triple-$\mathbf{k}$ magnetic structure with $C_3$ symmetry, deviating from the expected multi-domain zigzag structure (Yao \emph{et al.}, 2023). Measuring spin-wave spectra is another valuable approach for distinguishing multi-$\mathbf{k}$ features. The distinct shapes of magnetic Brillouin zones result in different symmetries of the magnetic excitation spectra, allowing to identify and characterize these structures. By using this approach the triple-$\mathbf{k}$ state with a vortex spin structure was further confirmed in the hexagonal Na$_{2}$Co$_{2}$TeO$_{6}$ compound (Chen \emph{et al.}, 2021; Yao \emph{et al.}, 2022; Kr\"{u}ger \emph{et al.}, 2023).

In this section we will provide the crystallographic description for such double or triple-$\mathbf{k}$ magnetic structures that can emerge from zigzag-type structures. Magnetic structures models have been constructed using the magnetic symmetry tools available at the Bilbao Crystallographic Server (BCS) (Perez-Mato \emph{et al.}, 2015).

\subsection{Double-$\mathbf{k}$ magnetic structures in the monoclinic Na$_{3}$Co$_{2}$SbO$_{6}$ and $m$-Na$_{2}$Co$_{2}$TeO$_{6}$ systems}

For the monoclinic Na$_{3}$Co$_{2}$SbO$_{6}$ and Na$_{2}$Co$_{2}$TeO$_{6}$ compounds, the 1-$\mathbf{k}$ zigzag-type magnetic structures can be defined by either $\pm$($\nicefrac{1}{2}$, $\nicefrac{1}{2}$, 0) and $\pm$($\nicefrac{1}{2}$, -$\nicefrac{1}{2}$, 0) wave vectors. Each possibility gives rise to coexisting $\mathbf{k}$-domains, related by a 180$^{\circ}$ rotation about the $b$-axis. For a 2-$\mathbf{k}$ scenario, the configuration of the magnetic moment is derived by superimposing the two zigzag components. The resulting vector depends on the actual orientation of the magnetic moment in the zigzag chains, relative to the $\mathbf{k}$-vector. It’s worth noting that the magnetic space group $P_S\bar{1}$, which describes the 1-$\mathbf{k}$ structures, allows the orientation of the magnetic moment along any crystallographic axis. For Na$_{3}$Co$_{2}$SbO$_{6}$, one proposed model suggested the alignment of the magnetic moments along the $b$-axis (Yan \emph{et al.}, 2019). A more recent model proposed that  the moments are aligned within the $ab$ plane, perpendicular to the wave vector (Li \emph{et al.}, 2022; Gu \emph{et al.}, 2024). In the former model, the vector superposition in 2-$\mathbf{k}$ structure results in two magnetic sublattices where half of the moments align along the $b$-axis, while the other half do not have a static ordered moment. 
Conversely, a pair of single-$\mathbf{k}$ structures with moments perpendicular to $\mathbf{k}$ will produce components that are nearly 60$^{\circ}$ or 120$^{\circ}$ apart, due to proximity to the $C_3$-symmetry. In this case, the 2-$\mathbf{k}$ structure will consist of two orthogonal sublattices with a ratio between moment amplitudes close to $\sqrt{3}$ : 1.

A crystallographic representation of this 2-$\mathbf{k}$ magnetic structure was obtained using the k-SUBGROUPSMAG program hosted at the BCS (Perez-Mato \emph{et al.}, 2015). The magnetic space group $C_a 2/m$ (\# 12.64) was utilized in the (2a, 2b, c) magnetic unit cell, which divides the Co position into two non-equivalent sites with distinct symmetry constraints. One site (Co1) confines the magnetic moment to the $ac$-plane, while the second site (Co2) aligns the moment parallel to $b$-axis. To evaluate the moment magnitudes, we used as the starting point the $b$-component (0.9~$\mu_B$) reported by Yan \emph{et al.} (2019), and added an $a$-projection ($\approx$0.49~$\mu_B$) to ensure orthogonality to the $\mathbf{k}$-vector (as proposed by Gu \emph{et al.}). The moments of the 2-$\mathbf{k}$ model were refined using the Fullprof program (Rodríguez-Carvajal, 1993) against the simulated data from the single-$\mathbf{k}$ model. The obtained moments for the two Co sublatices are m$_{Co1}$ = (0.69, 0, 0) $\mu_B$ and  m$_{Co2}$ = (0, 1.27, 0) $\mu_B$.

\begin{figure}
\includegraphics[width=6in]{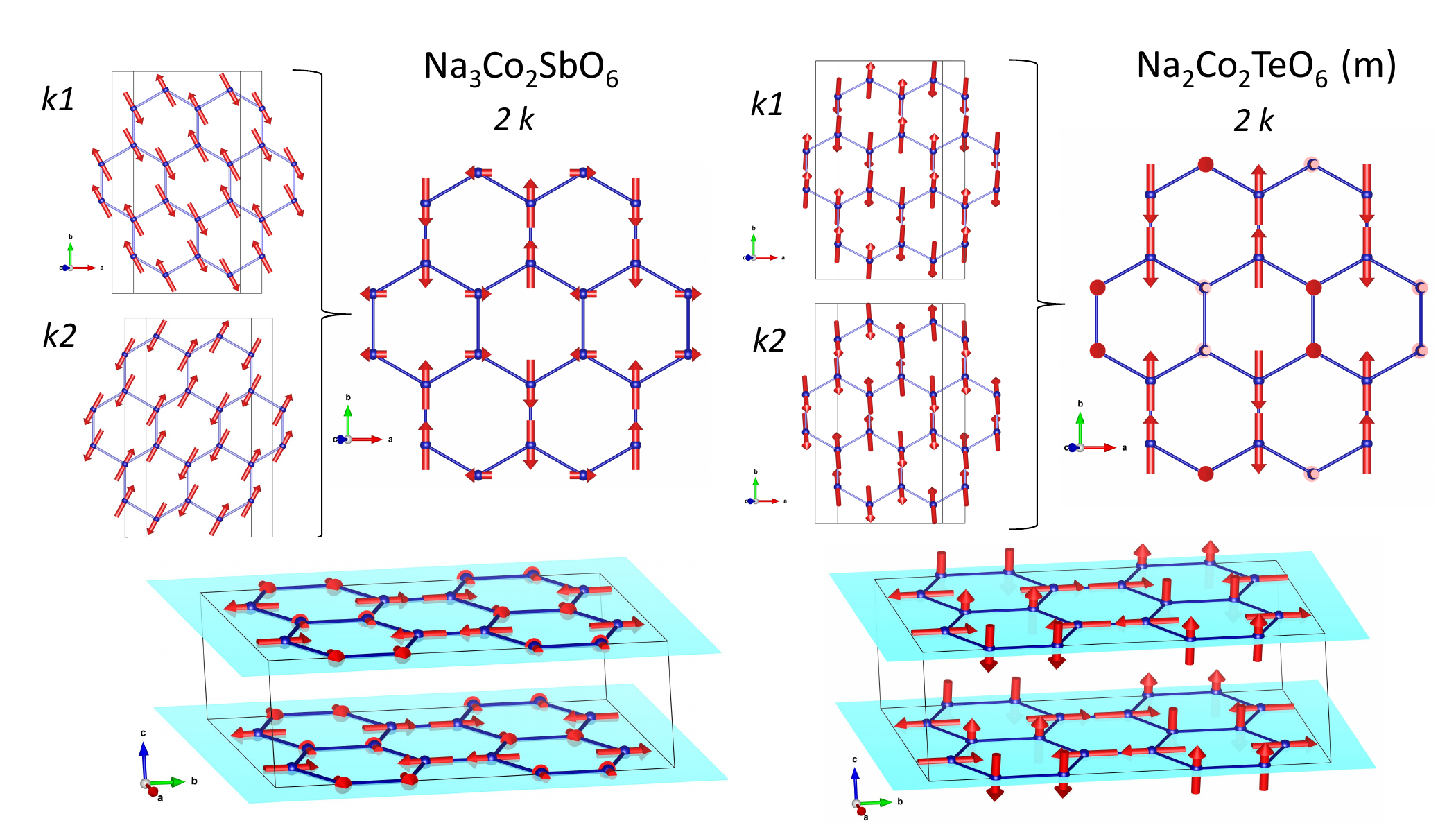}
\caption{Spin patterns of two types of zigzag domains and the 2-$\mathbf{k}$ structures for Na$_{3}$Co$_{2}$SbO$_{6}$ and monoclinic Na$_{2}$Co$_{2}$TeO$_{6}$.}
\label{fig:twok}
\end{figure}

The 1-$\mathbf{k}$ zigzag magnetic structure of the monoclinic polymorph of Na$_{2}$Co$_{2}$TeO$_{6}$ was shown to exhibit an out-of-plane magnetic component (Dufault \emph{et al.}, 2023). Consequently, its corresponding 2-$\mathbf{k}$ variant will represent a more generalized case, with a $c$-component on half of the Co moments. We computed the moments for a potential 2-$\mathbf{k}$ magnetic structure, described by $C_a 2/m$, to match the diffraction intensities simulated from the single-$\mathbf{k}$ structure. The magnetic moments on the two non-equivalent Co positions are found to be m$_{Co1}$ = (-0.69, 0, -1.69)~$\mu_B$ and m$_{Co2}$ = (0.0, 2.1, 0.0) $\mu_B$. The in-plane spin patterns of two types of zigzag domains and the 2-$\mathbf{k}$ structures are illustrated in Fig.~\ref{fig:twok}, for both Na$_{3}$Co$_{2}$SbO$_{6}$ and $m$-Na$_{2}$Co$_{2}$SbO$_{6}$. The complete descriptions of the 2-$\mathbf{k}$ models are given in Table~\ref{table_multik}. The magnetic-CIF files are available as part of the supplementary materials.

\subsection{Triple-$\mathbf{k}$ magnetic structures in the hexagonal Na$_{2}$Co$_{2}$TeO$_{6}$ systems}

\begin{figure}
\caption{The spin patterns of three distinct zigzag domains and the alternative 3-$\mathbf{k}$ magnetic structure, within the hexagonal Na$_{2}$Co$_{2}$TeO$_{6}$.}
\includegraphics[width=4in]{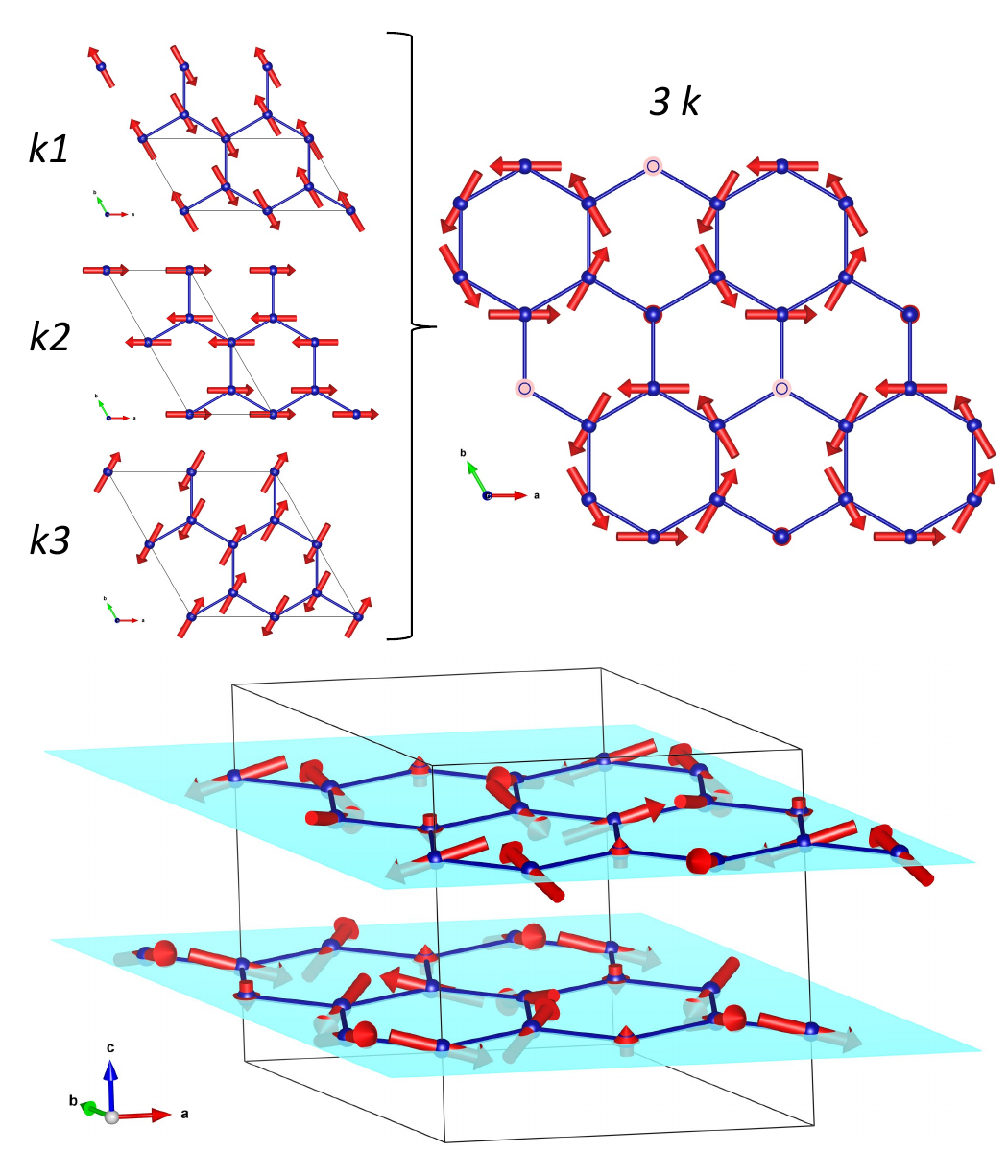}
\label{fig:threek}
\end{figure}

The triple-$\mathbf{k}$  state in the hexagonal Na$_{2}$Co$_{2}$TeO$_{6}$ is formed by combining three $C_3$-related single-$\mathbf{k}$ zigzag components, each with a different orientation. In the 1-$\mathbf{k}$ zigzag model, the ferromagnetic chains and the moments align perpendicular to the wave vector, with the magnetic moments positioned in the $bc$ plane and a minor ($\approx$0.5$\mu_B$) out-of-plane component (Samarakoon \emph{et al.}, 2021). By superposing contributions of the three wave vectors, ($\nicefrac{1}{2}$, 0, 0), (0, -$\nicefrac{1}{2}$, 0) and (-$\nicefrac{1}{2}$, $\nicefrac{1}{2}$, 0) the magnetic moments form spin vertices, where canted moments at the corner of hexagons are rotated by 60$^{\circ}$ and alternate in their canting direction. The vertices are separated by sites with the alternating $c$-axis moments, as depicted in Fig.~\ref{fig:threek}. From a crystallographic perspective, this structure can be described using the magnetic space group $P 6_32'2'$ (\# 182.183) in a magnetic supercell that is doubled along both $a$ and $b$ axes (2a, 2b, c). In this symmetry, each of the two non-identical Co sites of the parent structure is divided into two additional magnetic sites (with multiplicities 6 and 2), leading to four magnetic sublattices. Two of the sites, which represent 1/3 of Co sites, confine the magnetic moments to the $c$ direction. The other two sites (2/3 of total sites) allow components in- and out-of-plane with a constrained 60$^{\circ}$ rotation within the $ab$ plane. Using the moment values of the 1-$\mathbf{k}$ model reported by Samarakoon \emph{et al.}, we derived the actual moment values for the alternative triple-$\mathbf{k}$. As was done previously for the zigzag model, we assigned the same moment amplitudes to Co sites with similar magnetic symmetries. The obtained moments components in the triple-$\mathbf{k}$ model were: m$_{Co11}$ = (2.31, 0, -0.5)~$\mu_B$,  m$_{Co12}$ = (0, 0, 0.5)~$\mu_B$, m$_{Co21}$ = (2.31, 2.31, 0.5)~$\mu_B$ and m$_{Co22}$ = (0, 0, -0.5)~$\mu_B$. Further description of this 3-$\mathbf{k}$ model is provided in Table~\ref{table_multik}. The magnetic-CIF file is provided as supplementary material.

\section{Concluding remarks}

Layered materials with a honeycomb lattice are incredibly versatile and can host a variety of magnetic ground states. These states are dictated by the relative ratio and signs of the nearest-neighbor ($J_1$), next-nearest neighbor ($J_2$) or third-nearest neighbor ($J_3$) couplings. In this review article, our focus has been on materials that exhibit alternating FM-zigzag chains, as they appear to be, aside from some spiral orders, a dominant magnetic ground state in Kitaev-type candidate materials. This type of structure can be stabilized by a combination of a FM $J_1$ and AFM $J_3$. They are interesting in the context of Kitaev physics due to their prevalence in material candidates proximate to the elusive quantum spin-liquid state. The studies of zigzag-type structures in materials have also led to the development of extended Kitaev Hamiltonians, which include additional symmetry-allowed perturbing interactions, necessary to describe real materials. These lead to an alternative route for stabilizing the zigzag structure at a nearest neighbor level via off-diagonal exchange terms $\Gamma$, $\Gamma'$.

In this work, we have highlighted the stability of the zigzag structure across a wide range of parent crystal symmetries ($C2/m$, $C2/c$, $P6_322$, $P6_3/mcm$, $R\bar{3}$) in representative compounds. These compounds contain 5$d$, 4$d$ or 3$d$ transition-metal magnetic ion species (Ir$^{4+}$, Ru$^{3+}$, Co$^{2+}$) with an effective spin 1/2. The Kitaev model, originally formulated for spin-1/2 systems, has been a cornerstone in the study of quantum spin liquids due to its exactly solvable nature. Extending the Kitaev model to larger spins $S\geq$1 is seen as an exciting opportunity to unveil new physics and broaden the scope of Kitaev materials. Therefore, we have also included in our review the Ni$^{2+}$ based honeycombs, as recent experimental studies on these $S$ = 1 materials have provided evidence for the presence of bond-anisotropic Kitaev interactions. 

The crystallographic details of the zigzag structure observed in several classes of materials have been reviewed and tabulated. Notably, the magnetic moments in Na$_2$IrO$_3$ and cobaltates in the $A_{2}$Co$_{2}M'$O$_{6}$ series, primarily reside within the honeycomb plane or display minor canting angles. The out-of-plane canting is relevant in the context of the anisotropic bond-directional Kitaev interaction. For some of these materials, the presence of out-of-plane components is still awaiting experimental confirmations. In contrast, the Ni compounds exhibit an easy axis anisotropy that tends to orient the moment along the $c$-axis. Reversed anisotropies are noted in the K$M$AsO$_4$ compounds, where the Ni moments are predominantly in the $ab$-plane and Co moments are nearly along the $c$-axis. The magnetic moments also lie in, or close to, the honeycomb plane in BaNi$_{2}$(AsO$_{4}$)$_2$ and in the van-der-Waals materials MP$X_3$. In RuCl$_3$ the canting angle, reported to be about 48$^{\circ}$, is more pronounced than in any other compounds. This canting demonstrates a strong dependence on the chemical composition, with a 10\% substitution of Ru by Rh or Ir leading to a 15$^{\circ}$ angle.

The slightly distorted honeycomb lattices of the monoclinic compounds exhibit two distinct types of propagation vectors: ($\nicefrac{1}{2}$,~$\nicefrac{1}{2}$,~0) or (0,~1,~0). The direction of the wave vector defines the direction of the ferromagnetic chain, and thus the nature of the bonds involved in the magnetic coupling. While a $\mathbf{k}$ = (0,~1,~0) or (1,~0,~0) implies the formation of zigzag chains with uniform bond length, the ($\nicefrac{1}{2}$,~$\nicefrac{1}{2}$,~0) produces ferromagnetic chains in the diagonal direction that involve alternating bonds. The latter ordering vector is also susceptible to forming multi-$\mathbf{k}$ structures. Variations in the ordering wave vector are also encountered in the hexagonal or rhombohedral systems that are intrinsically susceptible to stacking faults. In such systems, the magnetic propagation vectors can involve a doubling of the lattice along the $c$-direction, which is very sensitive to the sample composition. In particular, the alkali metal distribution amidst the honeycomb layers can be markedly disordered and inhomogeneous, and that can contribute to the stacking sequence of the magnetic layers. Mixed propagation vectors, suggesting stacking disorder, was observed in Na$_{2}$Ni$_{2}$TeO$_{6}$ and KCoAsO$_4$ compounds.

Inspired by recent elastic and inelastic neutron-scattering experiments on high-quality single crystals that uncovered the existence of multi-$\mathbf{k}$ magnetic structures in Na$_{3}$Co$_{2}$SbO$_{6}$ and Na$_{2}$Co$_{2}$TeO$_{6}$ compounds, we have provided detailed crystallographic information to better define these structures. Such structures suggest the presence of non-trivial, symmetry-allowed, non-bilinear exchange perturbations. As a result, we anticipate a new course in probing and examining the multi-$\mathbf{k}$ variants will emerge. There are many more honeycomb compounds with zigzag-type order that are yet to be discovered or are awaiting confirmation of their magnetic structures by neutron scattering experiments. It is our hope that this review will stimulate further interest and research efforts in exploring these fascinating systems.




\ack{This work used resources at the Spallation Neutron Source, DOE Office of Science User Facilities operated by the Oak Ridge National Laboratory. This manuscript has been authored by UT-Battelle, LLC under Contract No. DE-AC05-00OR22725 with the U.S. Department of Energy. The United States Government retains and the publisher, by accepting the article for publication, acknowledges that the United States Government retains a non-exclusive, paid-up, irrevocable, world-wide license to publish or reproduce the published form of this manuscript, or allow others to do so, for United States Government purposes. The Department of Energy will provide public access to these results of federally sponsored research in accordance with the DOE Public Access Plan (http://energy.gov/downloads/doe-public-access-plan)}

\appendix
\section{Tabulated crystallographic description of the discussed magnetic structures}

\begin{table}
\caption{Crystallographic description of zigzag magnetic structures observed in monoclinic honeycomb systems (I).}
\scriptsize
\centering
\label{table_mono1}
\begin{tabular}{p{2.2cm}K{3.2cm}K{3.2cm}K{3.2cm}} \hline
Compound & Na$_{2}$IrO$_{3}$  & Na$_{3}$Co$_{2}$SbO$_{6}$ & Na$_{2}$Co$_{2}$TeO$_{6}$ \\ [0.75ex]\hline \hline
MAGNDATA\# \par References & 1.10 \par (Ye, 2017) &  1.788 \par (Yan, 2019) &  1.259 \par (Dufault, 2023) \\ \hline
Parent SG & $C 2/m$ & \multicolumn{2}{K{6cm}}{$C 2/m$} \\ \hline
$\mathbf{k}$-vector & (0, 1, $\nicefrac{1}{2}$ ) & \multicolumn{2}{K{6cm}}{($\nicefrac{1}{2}$, $\nicefrac{1}{2}$, 0 )} \\ \hline
Transf. from parent to magnetic cell & (a$_p$, b$_p$, 2c$_p$) & \multicolumn{2}{K{6cm}}{(2a$_p$, 2b$_p$, c$_p$)} \\ \hline
MSG symbol \par (UNI notation) & $C_c 2/m $ \par ($C2/m.1'_c[C2/m]$) & \multicolumn{2}{K{6cm}}{$P_S\text{-}1$ \par ($P\text{-}1.1'_c[P\text{-}1]$)} \\ \hline
MSG \# & 12.63 & \multicolumn{2}{K{6cm}}{2.7} \\ \hline
Transf. to MSG standard setting & a-c, b, c; 0, 0, $\nicefrac{1}{4}$ & \multicolumn{2}{K{6cm}}{c, -$\nicefrac{1}{4}$a+$\nicefrac{1}{4}$b, -$\nicefrac{1}{2}$a-$\nicefrac{1}{2}$b; $\nicefrac{1}{8}$, $\nicefrac{1}{8}$, 0} \\ \hline
Magnetic PG & \textit{2/m1'} & \multicolumn{2}{K{6cm}}{\textit{-1.1'}} \\ \hline
a, b, c ({\AA}) \par $\alpha$, $\beta$, $\gamma$ &  5.32, 9.215, 11.07 \par 90$^{\circ}$, 108.67$^{\circ}$, 90$^{\circ}$ & 10.741, 18.578, 5.653 \par 90$^{\circ}$, 108.56$^{\circ}$, 90$^{\circ}$ & 10.656, 18.387, 5.769 \par 90$^{\circ}$, 108.95$^{\circ}$, 90$^{\circ}$ \\ \hline
MSG symmetry operations & x,y,z,+1 \par -x,y,-z+$\nicefrac{1}{2}$,+1 \par -x,-y,-z+$\nicefrac{1}{2}$,+1 \par x,-y,z,+1 & \multicolumn{2}{K{6cm}}{x,y,z,+1 \par -x+$\nicefrac{1}{4}$,-y+$\nicefrac{1}{4}$,-z,+1} \\ \hline
MSG centering operations & x,y,z,+1 \par x+$\nicefrac{1}{2}$,y+$\nicefrac{1}{2}$,z+$\nicefrac{1}{2}$,+1 \par x,y,z+$\nicefrac{1}{2}$,-1 \par x+$\nicefrac{1}{2}$,y+$\nicefrac{1}{2}$,z,-1 & \multicolumn{2}{K{6cm}}{x,y,z,+1 \par x+$\nicefrac{1}{4}$,y+$\nicefrac{3}{4}$,z,+1 \par x+$\nicefrac{1}{2}$,y+$\nicefrac{1}{2}$,z,+1 \par x+$\nicefrac{3}{4}$,y+$\nicefrac{1}{4}$,z,+1 \par x+$\nicefrac{3}{4}$,y+$\nicefrac{3}{4}$,z,-1 \par x,y+$\nicefrac{1}{2}$,z,-1\par x+$\nicefrac{1}{4}$,y+$\nicefrac{1}{4}$,z,-1 \par x+$\nicefrac{1}{2}$,y,z,-1} \\ \hline
Positions of magnetic atoms & Ir1  0 0.333 0 \par Ir2  0 0 0 & Co 0  0.333 0 & Co 0 0.337 0  \\ \hline
Non-magnetic atoms & Na1 0 0 0.333 \par Na2 0 0 0  \par Na3 0 0.836 0.25 \par Na4 0 0.5 0.25 \par O1 0.26 0.329 0.396 \par O2 0.27 0 0.396 & Sb 0 0 0 \par Na1 0 0.25 0.5 \par Na2 0 0.413 0.5 \par O11 0.136  0.17 0.794 \par  O12 0.863 0.17 0.205 \par O2 O 0.375 0 0.20 & Te  0 0 0 \par Na1 0 0.25 0.5 \par Na2 0 0.413 0.5 \par O11 0.14 0.172 0.80 \par O12 0.86 0.172 0.197 \par O2 0.381 0 0.192  \\ \hline
magnetic moments ($\mu_{B}$), sym. constraints  and total magnitudes & Ir1 (0.22 ,0 ,0) \par ($m_x$,0,$m_z$) 0.22 \par Ir2 (0, 0, 0) \par (0, 0, 0) 0 & Co (0, 0.9, 0) \par ($m_x$,$m_y$,$m_z$)  0.9 & Co (0.48, 1.5, 1.2) \par ($m_x$,$m_y$,$m_z$) 1.83 \\ \hline
\end{tabular}
\end{table}

\begin{table}
\caption{Crystallographic description of zigzag magnetic structures observed in monoclinic honeycomb systems (II).}
\scriptsize
\centering
\label{table_mono2}
\begin{tabular}{p{2.2cm}K{3.2cm}K{3.2cm}K{3.2cm}}\hline
Compound & Li$_{3}$Ni$_{2}$SbO$_{6}$ & Na$_{3}$Ni$_{2}$SbO$_{6}$ & Na$_{3}$Ni$_{2}$BiO$_{6}$ \\ [0.75ex] \hline \hline
MAGNDATA\# \par References &  1.373 \par (Kurbakov, 2017) &  - \par (this work) &  1.789 \par (Seibel, 2013)  \\ \hline
Parent SG & \multicolumn{2}{K{6cm}}{$C 2/m$} & $C 2/m$  \\ \hline
$\mathbf{k}$-vector & \multicolumn{2}{K{6cm}}{($\nicefrac{1}{2}$, $\nicefrac{1}{2}$, 0 )} & (0, 1, 0)\\ \hline
Transf. from parent to magnetic basis & \multicolumn{2}{K{6cm}}{(2a$_p$, 2b$_p$, c$_p$)} & (a$_p$, b$_p$, c$_p$) \\ \hline
MSG symbol \par (UNI) & \multicolumn{2}{K{6cm}}{$P_S\text{-}1$ \par ($P\text{-}1.1'_c[P\text{-}1]$)} & $P_C 2_1/m $ \par ($P2_1/m.1'_C[C2/m]$) \\ \hline
MSG \# & \multicolumn{2}{K{6cm}}{2.7} & 11.57\\ \hline
Transf. to standard MSG setting & \multicolumn{2}{K{6cm}}{c, -$\nicefrac{1}{4}$a+$\nicefrac{1}{4}$b, -$\nicefrac{1}{2}$a-$\nicefrac{1}{2}$b; $\nicefrac{1}{8}$, $\nicefrac{1}{8}$, 0 } & a, b, c; $\nicefrac{1}{4}$, $\nicefrac{1}{4}$, 0 \\ \hline
Magnetic PG & \multicolumn{2}{K{6cm}}{\textit{-1.1'}} & \textit{2/m1'} \\ \hline
a, b, c ({\AA}) \par $\alpha$, $\beta$, $\gamma$ & 10.368, 17.942, 5.16 \par 90$^{\circ}$, 109.7$^{\circ}$, 90$^{\circ}$ & 10.488, 18.331, 5.581 \par 90$^{\circ}$, 108.27$^{\circ}$, 90$^{\circ}$ & 5.399, 9.352, 5.68 \par 90$^{\circ}$, 108.56$^{\circ}$, 90$^{\circ}$\\ \hline
MSG symmetry operations & \multicolumn{2}{K{6cm}}{x,y,z,+1 \par -x+$\nicefrac{1}{4}$,-y+$\nicefrac{1}{4}$,-z,+1} & x,y,z,+1 \par  -x+$\nicefrac{1}{2}$,y+$\nicefrac{1}{2}$,-z,+1 \par -x+$\nicefrac{1}{2}$,-y+$\nicefrac{1}{2}$,-z,+1 \par x,-y,z,+1 \\ \hline
MSG symmetry centering operations & \multicolumn{2}{K{6cm}}{x,y,z,+1 \par x+$\nicefrac{1}{4}$,y+$\nicefrac{3}{4}$,z,+1 \par x+$\nicefrac{1}{2}$,y+$\nicefrac{1}{2}$,z,+1 \par x+$\nicefrac{3}{4}$,y+$\nicefrac{1}{4}$,z,+1 \par x+$\nicefrac{3}{4}$,y+$\nicefrac{3}{4}$,z,-1 \par x,y+$\nicefrac{1}{2}$,z,-1\par x+$\nicefrac{1}{4}$,y+$\nicefrac{1}{4}$,z,-1 \par x+$\nicefrac{1}{2}$,y,z,-1} & x,y,z,+1 \par x+$\nicefrac{1}{2}$,y+$\nicefrac{1}{2}$,z,-1 \\ \hline
Positions of magnetic atoms & Ni 0 0.1666 0 & Ni 0 0.333 0 & Ni 0 0.666 0 \\ \hline
Non-magnetic atoms & Sb 0 0 0 \par O1 0.379 0 0.225 \par O21 0.118 0.08 0.233 \par O22 0.88 0.078 0.766 \par Li1 0 0.08 0.5 \par Li2 0 0.25 0.5
 & Sb 0 0 0 \par O11 0.136 0.175 0.79 \par O12 0.863 0.175 0.21 \par O2 0.378 0 0.205 \par Na1 0 0.25 0.5 \par Na2 0 0.419 0.5 & Bi1 0 0 0 \par O1 0.226 0.84 0.20 \par O2 0.748 0 0.213 \par Na1 0 0.5 0.5 \par Na2 0 0.836 0.5 \\ \hline
Refined moments components ($\mu_{B}$), sym. constraints and total magnitudes & Ni (0.58, 0, 1.72) \par ($m_x$,$m_y$,$m_z$) 1.62 & Ni (0.68, -0.05, 1.9) ($m_x$,$m_y$,$m_z$) 1.8 & Ni (0.71, 0.0, 2.32) \par ($m_x$,0,$m_z$) 2.2 \\ \hline
\end{tabular}
\end{table}

\begin{table}
\caption{Crystallographic description of zigzag magnetic structures observed in monoclinic honeycomb systems (III).}
\scriptsize
\label{table_mono3}
\begin{tabular}{p{1.9cm}K{2.7cm}K{2.7cm}K{2.7cm}K{2.7cm}}\hline
Compound  & CoPS$_{3}$  & NiPS$_{3}$ & Cu$_{3}$Co$_{2}$SbO$_{6}$ & Cu$_{3}$Ni$_{2}$SbO$_{6}$ \\ [0.75ex] \hline \hline
MAGNDATA\# \par Reference &  1.264 \par (Wildes, 2017) & 1.231 \par(Wildes, 2015) & 1.258 \par (Roudebush, 2018) &  1.259 \par (Roudebush, 2018) \\ \hline
Parent SG & $C 2/m$ & $C 2/m$ & $C 2/c$ & $C 2/c$ \\ \hline
$\mathbf{k}$-vector & (0, 1, 0 ) & (0, 1, 0 ) & (1, 0, 0 ) & (1, 0, 0 ) \\ \hline
Transf. from parent to magnetic basis & (a$_p$, b$_p$, c$_p$) & (a$_p$, b$_p$, c$_p$) & (a$_p$, b$_p$, c$_p$) & (a$_p$, b$_p$, c$_p$) \\ \hline
MSG symbol \par (UNI) & $P_C 2_1/m$ \par ($P2_1/m.1'_C[C2/m]$) & $P_S 1$ \par ($P1.1'_c[P1]$) & $P_C 2_1/c$ \par ($P2_1/c.1'_C[C2/c]$) & $P_C 2/c$ \par ($P2/c.1'_C[C2/c]$) \\ \hline
MSG number & 11.57 & 1.3 & 14.84 & 13.74 \\ \hline
Trans. to standard setting & a, b, c; $\nicefrac{1}{4}$, $\nicefrac{1}{4}$, 0 & -c, b, a+b; 0, 0, 0 & a, b, a+c; 0, 0, 0 & a, b, c; 0, 0, 0 \\ \hline
Magnetic PG & \textit{2/m1'} & \textit{1.1'} & \textit{2/m1'} & \textit{2/m1'} \\ \hline
a, b, c ({\AA}) \par $\alpha$, $\beta$, $\gamma$ & 5.90, 10.22, 6.658 \par 90$^{\circ}$, 107.17$^{\circ}$, 90$^{\circ}$ & 5.811, 10.064, 6.594 \par 90$^{\circ}$, 106.997$^{\circ}$, 90$^{\circ}$ & 9.327, 5.385, 11.911 \par 90$^{\circ}$, 105.13$^{\circ}$, 90$^{\circ}$ & 9.192, 5.309, 11.915 \par 90$^{\circ}$, 104.88$^{\circ}$, 90$^{\circ}$ \\ \hline
MSG symmetry operations & x,y,z,+1 \par -x+1/2,y+$\nicefrac{1}{2}$,-z,+1 \par -x+$\nicefrac{1}{2}$,-y+$\nicefrac{1}{2}$,-z,+1 \par x,-y,z,+1 & x,y,z,+1 & x,y,z,+1 \par -x+$\nicefrac{1}{2}$,y+$\nicefrac{1}{2}$,-z+$\nicefrac{1}{2}$,+1 \par -x,-y,-z,+1 \par x+$\nicefrac{1}{2}$,-y+$\nicefrac{1}{2}$,z+$\nicefrac{1}{2}$,+1 & x,y,z,+1 \par -x,y,-z+$\nicefrac{1}{2}$,+1 \par -x,-y,-z,+1 \par x,-y,z+$\nicefrac{1}{2}$,+1 \\ \hline
MSG symmetry centering operations & x,y,z,+1 \par x+$\nicefrac{1}{2}$,y+$\nicefrac{1}{2}$,z,-1 & x,y,z,+1 \par x+$\nicefrac{1}{2}$,y+$\nicefrac{1}{2}$,z,-1 & x,y,z,+1 \par x+$\nicefrac{1}{2}$,y+$\nicefrac{1}{2}$,z,-1 & x,y,z,+1 \par x+$\nicefrac{1}{2}$,y+$\nicefrac{1}{2}$,z,-1 \\ \hline
Positions of magnetic atoms & Co 0 0.331 0 & Ni11 0 0.333 0 \par Ni12 0 0.666 0 \par Ni2 0 0 0 & Co1 0.08 0.249 0.499 & Ni1 0.08 0.249 0.499 \\ \hline
Non-magnetic atoms & P 0.057 0 0.169 \par S 0.746 0 0.244 \par S 0.252 0.17 0.247& P11 0.059 0 0.17 \par P12 0.44 0.5 0.83 \par P21 0.059 0.333 0.17 \par P22 0.44 0.166 0.83 \par P23 0.059 0.666 0.17 \par P24 0.44 0.833 0.83 \par S11 0.242 0.5 0.244 \par S12 0.257 0 0.755 \par S21 0.249 0.33 0.756 \par S22 0.25 0.17 0.243 \par S23 0.249 0.67 0.756 \par S24 0.25 0.83 0.243 & Cu1 0.823 0.09 0.249 \par Cu2 0 0.587 0.25 \par O1 0.624 0.593 0.09 \par O2 0.72 0.57 0.41 \par O3 0.442 0.085 0.09 \par Sb 0.25 0.25 0 & Cu1 0.825 0.09 0.249 \par Cu2 0 0.585 0.25 \par O1 0.62 0.588 0.09 \par O2 0.723 0.576 0.41 \par O3 0.44 0.08 0.088 \par Sb 0.25 0.25 0 \\ \hline
Refined moments components ($\mu_{B}$), sym. constraints and total moment & Co (3.13, 0, -0.6) \par ($m_x$,0,$m_z$) 3.36 & Ni11 (0.94, 0, -0.26) \par ($m_x$,$m_y$,$m_z$) 1.05 \par Ni12 (-0.94, 0, 0.26) \par ($m_x$,$m_y$,$m_z$) 1.05 \par Ni2 (1.0, 0.5, 0) \par ($m_x$,$m_y$,$m_z$) 1.12 & Co1 (0, 2.4, 0) \par ($m_x$,$m_y$,$m_z$) 2.4 & Ni1 (0.44, 0, 1.49) \par ($m_x$,$m_y$,$m_z$) 1.44 \\ \hline
\end{tabular}
\end{table}

\begin{table}
\caption{Crystallographic description of zigzag magnetic structures observed in hexagonal honeycomb systems.}
\scriptsize
\centering
\label{table_hex}
\begin{tabular}{p{2.2cm}K{3.1cm}K{3.1cm}K{3.1cm}}\hline
Compound & Na$_{2}$Co$_{2}$TeO$_{6}$ & Na$_{2}$Ni$_{2}$TeO$_{6}$ & Na$_{2}$Ni$_{2}$TeO$_{6}$ \\ [0.75ex] \hline \hline
MAGNDATA\# \par References & 1.645 \par (Samarakoon, 2021) &  1.646 \par (Samarakoon, 2021) &  1.258 \par (Kurbakov, 2020) \\ \hline
Parent SG & $P 6_3 2 2$ & \multicolumn{2}{K{6cm}}{$P 6_3/m c m$} \\ \hline
$\mathbf{k}$-vector & ($\nicefrac{1}{2}$, 0, 0 ) & ($\nicefrac{1}{2}$, 0, $\nicefrac{1}{2}$ ) & ($\nicefrac{1}{2}$, 0, 0 ) \\ \hline
Transf. from parent to magnetic cell & (2a$_p$, b$_p$, c$_p$) & (2a$_p$, b$_p$, 2c$_p$) & (2a$_p$, b$_p$, c$_p$) \\ \hline
MSG symbol \par (UNI)  & $P_C 2_1 2_1 2_1$ \par ($P2_12_12_1.1'_C[C222_1]$) & $I_a m m 2$ \par ($Imm2.1'_a[Amm2]$) & $P_A n m a$ \par ($Pnma.1'_A[Amma]$) \\ \hline
MSG number & 19.29 & 44.234 & 62.453 \\ \hline
Transf. to MSG standard setting & a+b, b, c; $\nicefrac{1}{4}, $\nicefrac{1}{4}, $\nicefrac{1}{4}$ & c, a+b, b; 0, 0, $\nicefrac{1}{8}$ & c, a+b, b; $\nicefrac{1}{4}$, 0, 0 \\ \hline
Magnetic PG & \textit{2221'} & \textit{mm21'} & \textit{mmm1'} \\ \hline
a, b, c ({\AA}) \par $\alpha$, $\beta$, $\gamma$ &  10.517, 5.258, 11.17 \par 90$^{\circ}$, 90$^{\circ}$, 120$^{\circ}$ & 10.382, 5.191, 22.182 \par 90$^{\circ}$, 90$^{\circ}$, 120$^{\circ}$ & 10.403, 5.201, 11.172 \par 90$^{\circ}$, 90$^{\circ}$, 120$^{\circ}$ \\ \hline
MSG symmetry operations & x,y,z,+1 \par -x,-2x+y,-z,-1 \par x,2x-y,-z+$\nicefrac{1}{2}$,-1 \par  -x,-y,z+$\nicefrac{1}{2}$,+1 &
x,y,z,+1 \par -x,-2x+y,-z+$\nicefrac{1}{4}$,+1 \par x,y,-z+$\nicefrac{1}{4}$,+1 \par -x,-2x+y,z,+1 &
x,y,z,+1 \par -x,-y,z+$\nicefrac{1}{2}$,+1 \par x+$\nicefrac{1}{2}$,2x-y,-z,+1 \par -x+$\nicefrac{1}{2}$,-2x+y,-z+$\nicefrac{1}{2}$,+1 \par -x+$\nicefrac{1}{2}$,-y,-z,+1 \par x+$\nicefrac{1}{2}$,y,-z+$\nicefrac{1}{2}$,+1 \par -x,-2x+y,z,+1 \par x,2x-y,z+$\nicefrac{1}{2}$,+1 \\ \hline
MSG symmetry centering operations & x,y,z,+1 \par x+$\nicefrac{1}{2}$,y,z,-1 & x,y,z,+1 \par x+$\nicefrac{1}{2}$,y,z+$\nicefrac{1}{2}$,+1 \par x,y,z+$\nicefrac{1}{2}$,-1 \par x+$\nicefrac{1}{2}$,y,z,-1 & x,y,z,+1 \par x+$\nicefrac{1}{2}$,y,z,-1 \\ \hline
Positions of magnetic atoms & Co1 0 0 0.25 \par Co2 0.166 0.666 0.75 & Ni 0.167 0.667 0 & Ni 0.167 0.667 0 \\ \hline
Positions of non-magnetic atoms & Te 0.166 0.666 0.25 \par O11 0.32 0.974 0.345 \par O12 O 0.51 0.668 0.345 \par O13 0.165 0.356 0.345 \par Na21 0 0 0 \par Na31 0.35 0.066 0.99 \par Na32 0.967 0.636 0.99 \par Na33 0.68 0.298 0.99 & Te 0 0 0 \par O11 0.163 0 0.296 \par O12 0 0.327 0.296 \par O13 0.836 0. 0.546 \par O14 0 0.672 0.546 \par Na11 0.174 0 0.125 \par Na12 0 0.349 0.125 \par Na13 0.825 0 0.375 \par Na14 0 0.65 0.375 \par Na21 0.166 0.667 0.125 \par Na22 0.333 0.333 0.375 & Te 0 0 0 \par O11 0.159 0 0.594 \par O12 0 0.318 0.594 \par Na11 0.173 0 0.25 \par Na12 0 0.347 0.25 \par Na2 0.167 0.667 0.25 \par Na3 0 0 0.25 \\ \hline
Refined moments components ($\mu_{B}$), MSG constraints  and total magnitudes & Co1 (0, 2.07, 0.5) \par (0,$m_y$,$m_z$) 2.1 \par Co2 (0, -1.95, 0.5) \par (0,$m_y$,$m_z$) 2.0 & Ni (0, 0, 1.55) \par ($m_x$,$m_y$,$m_z$) 1.55 & Ni (0, 0.41, 1.67) \par (0,$m_y$,$m_z$) 1.72 \\ \hline
\end{tabular}
\end{table}

\begin{table}
\caption{Crystallographic description of zigzag magnetic structures observed in rhombohedral honeycomb systems.}
\scriptsize
\centering
\label{table_rho}
\begin{tabular}{p{1.8cm}K{2.7cm}K{2.7cm}K{2.8cm}K{2.8cm}}\hline
Compound & RuCl$_{3}$ & RuBr$_{3}$ & KNiAsO$_4$ & BaNi$_2$(AsO$_4$)$_2$ \\ [0.75ex] \hline \hline
MAGNDATA\# \par References & 1.787 \par (Park, 2024) & 1.786 \par (Imai, 2022) &  1.742 \par (Taddei, 2023) &  1.257 \par (Regnault, 1980) \\ \hline
Parent SG & $R\text{-}3$ & $R\text{-}3$ & $R\text{-}3$ & $R\text{-}3$ \\ \hline
$\mathbf{k}$-vector & \multicolumn{2}{K{6cm}}{(0, $\nicefrac{1}{2}$, 1)} & ($\nicefrac{3}{2}$, 0, 0) & ($\nicefrac{1}{2}$, 0, $\nicefrac{1}{2}$) \\ \hline
Transf. from parent to magn. cell & \multicolumn{2}{K{6cm}}{(a$_p$, 2b$_p$, c$_p$)} & (2a$_p$, b$_p$, c$_p$) & (2a$_p$, b$_p$, 2c$_p$) \\ \hline
MSG symbol \par (UNI label) & \multicolumn{2}{K{6cm}}{$P_S\text{-}1 $ \par ($P\text{-}1.1'_c[P\text{-}1]$)} & $P_S\text{-}1 $ \par ($P\text{-}1.1'_c[P\text{-}1]$) & $P_S\text{-}1$ \par ($P\text{-}1.1'_c[P\text{-}1]$) \\ \hline
MSG number & \multicolumn{2}{K{6cm}}{2.7} & 2.7 & 2.7 \\ \hline
Transf. to MSG standard setting & \multicolumn{2}{K{6cm}}{$\nicefrac{1}{3}$a+$\nicefrac{1}{3}$b-$\nicefrac{1}{3}$c, $\nicefrac{1}{3}$a+$\nicefrac{1}{3}$b+$\nicefrac{2}{3}$c, $\nicefrac{4}{3}$a+$\nicefrac{1}{3}$b+$\nicefrac{2}{3}$c;0,$\nicefrac{1}{4}$,0} & $\nicefrac{1}{3}$a+$\nicefrac{1}{3}$b+$\nicefrac{1}{3}$c, b, -$\nicefrac{1}{3}$a-$\nicefrac{4}{3}$b+$\nicefrac{2}{3}$c;$\nicefrac{1}{4}$,0,0 & $\nicefrac{1}{6}$a-$\nicefrac{1}{3}$b-$\nicefrac{1}{6}$c, $\nicefrac{1}{6}$a+$\nicefrac{2}{3}$b-$\nicefrac{1}{6}$c, $\nicefrac{2}{3}$a+$\nicefrac{2}{3}$b+$\nicefrac{1}{3}$c; 0,0,0 \\ \hline
Magnetic PG & \multicolumn{2}{K{6cm}}{\textit{-1.1'}} & \textit{-1.1'} & \textit{-1.1'} \\ \hline
a, b, c ({\AA}) \par $\alpha$, $\beta$, $\gamma$ &  5.973, 11.946, 16.93 \par 90$^{\circ}$, 90$^{\circ}$, 120$^{\circ}$ & 6.295, 12.591, 17.91 \par 90$^{\circ}$, 90$^{\circ}$, 120$^{\circ}$ & 9.973, 4.986, 28.619 \par 90$^{\circ}$, 90$^{\circ}$, 120$^{\circ}$ & 9.89, 4.945, 47.220 \par 90$^{\circ}$, 90$^{\circ}$, 120$^{\circ}$ \\ \hline
MSG symmetry operations & \multicolumn{2}{K{6cm}}{x,y,z,+1 \par -x,-y+$\nicefrac{1}{2}$,-z,+1} & x,y,z,+1 \par -x+$\nicefrac{1}{2}$,-y,-z,+1 &
x,y,z,+1 \par -x,-y,-z,+1 \\ \hline
MSG centering operations & \multicolumn{2}{K{6cm}}{x,y,z,+1 \par x+$\nicefrac{1}{3}$,y+$\nicefrac{1}{3}$,z+$\nicefrac{2}{3}$,+1 \par x+$\nicefrac{2}{3}$,y+$\nicefrac{2}{3}$,z+$\nicefrac{1}{3}$,+1 \par x+$\nicefrac{2}{3}$,y+$\nicefrac{1}{6}$,z+$\nicefrac{1}{3}$,-1 \par x,y+$\nicefrac{1}{2}$,z,-1 \par x+$\nicefrac{1}{3}$,y+$\nicefrac{5}{6}$,z+$\nicefrac{2}{3}$,-1} & x,y,z,+1 \par x+$\nicefrac{1}{3}$,y+$\nicefrac{1}{3}$,z+$\nicefrac{1}{3}$,+1 \par x+$\nicefrac{2}{3}$,y+$\nicefrac{2}{3}$,z+$\nicefrac{2}{3}$,+1 \par  x+$\nicefrac{5}{6}$,y+$\nicefrac{1}{3}$,z+$\nicefrac{1}{3}$,-1 \par  x+$\nicefrac{1}{6}$,y+$\nicefrac{2}{3}$,z+$\nicefrac{2}{3}$,-1 \par  x+$\nicefrac{1}{2}$,y,z,-1 & x,y,z,+1 \par x+$\nicefrac{1}{6}$,y+$\nicefrac{2}{3}$,z+$\nicefrac{5}{6}$,+1 \par x+$\nicefrac{1}{3}$,y+$\nicefrac{1}{3}$,z+$\nicefrac{2}{3}$,+1 \par x+$\nicefrac{1}{2}$,y,z+$\nicefrac{1}{2}$,+1 \par x+$\nicefrac{2}{3}$,y+$\nicefrac{2}{3}$,z+$\nicefrac{1}{3}$,+1 \par x+$\nicefrac{5}{6}$,y+$\nicefrac{1}{3}$,z+$\nicefrac{1}{6}$,+1 \par x+$\nicefrac{1}{3}$,y+$\nicefrac{1}{3}$,z+$\nicefrac{1}{6}$,-1 \par x+$\nicefrac{1}{2}$,y,z,-1 \par x+$\nicefrac{2}{3}$,y+$\nicefrac{2}{3}$,z+$\nicefrac{5}{6}$,-1 \par x+$\nicefrac{5}{6}$,y+$\nicefrac{1}{3}$,z+$\nicefrac{2}{3}$,-1 \par x,y,z+$\nicefrac{1}{3}$,-1 \par x+$\nicefrac{1}{6}$,y+$\nicefrac{2}{3}$,z+$\nicefrac{1}{3}$,-1 \par \\ \hline
Magnetic atoms pos. & Ru 0 0 0.333 & Ru 0 0 0.323 & Ni 0 0 0.1652 & Ni 0 0 0.083 \\ \hline
Non-magnetic atoms & Cl1 0.317 0.167 0.41 \par Cl2 0.665 0.991 0.41 \par Cl3 0.016 0.841 0.41 & Br1 0.329 0.49 0.08 \par Br2 0.016 0.67 0.08 \par Br3 0.654 0.835 0.08 & K 0 0  0.289 \par As 0 0 0.556 \par O11 0  0.345 0.124 \par O12 0.83 0.66 0.124 \par O13 0.17 0.99 0.124 \par O2 0 0 0.62 & Ba 0 0 0 \par As 0 0  0.213 \par O1 0 0  0.177 \par O21 0.164 0.328 0.06 \par O22 0.836 0 0.06 \par O23 0 0.672 0.06
\\ \hline
Magnetic moments ($\mu_{B}$), MSG constraints  and total magnitudes & Ru (0.49, 0, 0.54) \par ($m_x$,$m_y$,$m_z$) 0.73 & Ru (-0.3, 0.04, 0.66) \par ($m_x$,$m_y$,$m_z$) 0.74 & Ni (-0.2, 1.6, 0.7) ($m_x$,$m_y$,$m_z$) 1.8 & Ni (0, 2.0, 0) \par ($m_x$,$m_y$,$m_z$) 2.0 \\ \hline
\end{tabular}
\end{table}

\begin{table}
\caption{Crystallographic description of multi-$\mathbf{k}$ magnetic structures derived from zigzag structures.}
\scriptsize
\centering
\label{table_multik}
\begin{tabular}{p{2.5cm}K{3.2cm}K{3.2cm}K{3.2cm}}\hline
Compound & Na$_3$Co$_2$SbO$_6$ & Na$_2$Co$_2$TeO$_6$ & Na$_2$Co$_2$TeO$_6$ \\ [0.75ex] \hline \hline
Parent SG & \multicolumn{2}{K{6.4cm}}{$C 2/m$} & $P 6_3 2 2$ \\ \hline
$\mathbf{k}$-vector & \multicolumn{2}{K{6.4cm}}{2-$\mathbf{k}$: ($\nicefrac{1}{2}$, $\nicefrac{1}{2}$, 0) + \par ($\nicefrac{1}{2}$, -$\nicefrac{1}{2}$, 0)} & 3-$\mathbf{k}$: ($\nicefrac{1}{2}$, 0, 0) + \par (0, -$\nicefrac{1}{2}$, 0) + \par (-$\nicefrac{1}{2}$, $\nicefrac{1}{2}$, 0) \\ \hline
Transf. from parent to magnetic cell & \multicolumn{2}{K{6.4cm}}{(2a$_p$, 2b$_p$, c$_p$)} & (2a$_p$, 2b$_p$, c$_p$) \\ \hline
MSG symbol \par (UNI)& \multicolumn{2}{K{6.4cm}}{$C_a 2/m $ \par ($C2/m.1'_a[P2/m]$)} & $P 6_32'2'$ \par ($P 6_32'2'$) \\ \hline
MSG number & \multicolumn{2}{K{6.4cm}}{12.64} & 182.183 \\ \hline
Transf. to MSG standard setting & \multicolumn{2}{K{6cm}}{a, b, c; $\nicefrac{1}{4}$, 0, 0} & a, b, c; $\nicefrac{1}{4}$, 0, 0 \\ \hline
Magnetic PG & \multicolumn{2}{K{6.4cm}}{\textit{2/m1'}} & \textit{62'2'} \\ \hline
a, b, c ({\AA}) \par $\alpha$, $\beta$, $\gamma$ &  10.741, 18.578, 5.653 \par 90$^{\circ}$, 108.56$^{\circ}$, 90$^{\circ}$ & 10.656, 18.387, 5.769 \par 90$^{\circ}$, 108.95$^{\circ}$, 90$^{\circ}$ & 10.517, 10.517, 11.17 \par 90$^{\circ}$, 90$^{\circ}$, 120$^{\circ}$ \\ \hline
MSG symmetry operations & \multicolumn{2}{K{6.4cm}}{x,y,z,+1 \par  -x,-y,-z,-1 \par x,-y,z,+1 \par -x,y,-z,-1 }  &
x,y,z,+1 \par x-y,x+$\nicefrac{1}{2}$,z+$\nicefrac{1}{2}$,+1 \par -y+$\nicefrac{1}{2}$,x-y+$\nicefrac{1}{2}$,z,+1 \par -x,-y,z+$\nicefrac{1}{2}$,+1 \par -x+y,-x+$\nicefrac{1}{2}$,z,+1 \par y+$\nicefrac{1}{2}$,-x+y+$\nicefrac{1}{2}$,z+$\nicefrac{1}{2}$,+1 \par x-y,-y,-z,-1 \par y+$\nicefrac{1}{2}$,x+$\nicefrac{1}{2}$,-z,-1 \par -x,-x+y+$\nicefrac{1}{2}$,-z,-1 \par  x,x-y+$\nicefrac{1}{2}$,-z+$\nicefrac{1}{2}$,-1 \par  -x+y,y,-z+$\nicefrac{1}{2}$,-1 \par -y+$\nicefrac{1}{2}$,-x+$\nicefrac{1}{2}$,-z+$\nicefrac{1}{2}$,-1 \\ \hline
MSG centering operations & \multicolumn{2}{K{6.4cm}}{x,y,z,+1 \par x+$\nicefrac{1}{2}$,y+$\nicefrac{1}{2}$,z,+1 \par x,y+$\nicefrac{1}{2}$,z,-1 \par x+$\nicefrac{1}{2}$,y,z,-1} & x,y,z,+1 \\ \hline
Positions of magnetic atoms & Co1 0 0.333 0 \par Co2 0.25 0.583 0 & Co1 0 0.334 0 \par Co2 0.25 0.584 0 & Co11 0 0 0.25 \par Co12 0.5 0 0.25  \par Co21 0.333 0.166 0.25 \par  Co22 0.166 0.333 0.75 \\ \hline
Non-magnetic atoms & Sb1 0 0 0 \par Sb2 0.25 0.25 0 \par Na11 0 0.25 0.5 \par Na12 0.25 0 0.5 \par Na21 0 0.413 0.5 \par Na22 0.25 0.66 0.5 \par O11 0.136  0.17 0.794 \par O12 0.386 0.42 0.79 \par O21 0.375 0 0.204 \par O22 0.625 0.25 0.204 & Te1 0 0 0 \par Te2 0.25 0.25 0 \par O11 0.14 0.172 0.80 \par O12 0.390 0.422 0.80 \par O21 0.132 0 0.19 \par O22 0.382 0.25 0.19 \par Na11 0 0.164 0.5 \par Na12 0.25 0.414 0.5 \par Na21 0.0 0.25 0.5 \par Na22 0.25 0 0.5 & Te1 0.166 0.333 0.25 \par Te2 0.333 0.166 0.75 \par O11 0.324 0.49 0.345 \par O12 0.51 0.835 0.345 \par O13 0.164 0.675 0.345 \par O14 0.824 0.488 0.345 \par Na21 0 0 0 \par Na22 0.5 0 0 \par Na31 0.349 0.033 0.99 \par Na32 0.966 0.316 0.99 \par Na33 0.683 0.650 0.99 \par Na34 0.849 0.033 0.99 \\ \hline
Magnetic moments ($\mu_{B}$), sym. constraints and total magnitudes & Co1 (0.69, 0, 0) \par ($m_x$,0,$m_z$) 0.69 \par Co2 (0, 1.27, 0) \par (0,$m_y$,0) 1.27 & Co1 (-0.68, 0, -1.68) \par ($m_x$,0,$m_z$) 1.60 \par Co2 (0, 2.09, 0) \par (0,$m_y$,0)  2.09  & Co11 (2.31, 0, -0.5) ($m_x$,0,$m_z$)  2.36 \par Co12 (0, 0, 0.5) \par (0,0,$m_z$) 0.5 \par Co21 (2.31, 2.31, 0.5) \par ($m_x$,$m_x$,$m_z$) 2.36 \par Co22 (0, 0, -0.5) \par (0,0,$m_z$) 0.5  \\ \hline
\end{tabular}
\end{table}



\begin{references}
\reference{Anderson, Philip W. (1973), \emph{Materials Research Bulletin} \textbf{8.2}, 153-160.}
\reference{Balents, L. (2010), \emph{Nature} \textbf{464}, 199-208.} 
\reference{Balz, C., Lampen-Kelley, P., Banerjee, A., Yan, J., Lu, Z., Hu, X., Yadav, S.M., Takano, Y., Liu, Y., Tennant, D.A. and Lumsden, M.D. (2019), \emph{Phys. Rev. B} \textbf{100}, 060405.} %
\reference{Balz, C., Janssen, L., Lampen-Kelley, P., Banerjee, A., Liu, Y. H., Yan, J.-Q., Mandrus, D. G., Vojta, M., Nagler, S. E. (2021), \emph{Phys. Rev. B} \textbf{103}, 174417.} 
\reference{Banerjee, A. , Bridges, C. A., Yan, J.-Q. \textit{et al} (2016) \emph{Nat. Mater.} \textbf{15}, 733.}
\reference{Banerjee, A., Yan, J., Knolle, J., Bridges, C. A., Stone, M. B., Lumsden, M. D., Mandrus, D. G., Tennant, D. A., Moessner, R, Nagler, S. E. (2017) \emph{Science} \textbf{356} 1055.}
\reference{Banerjee, A., Lampen-Kelley, P., Knolle, J., Balz, C., Aczel, A. A., Winn, B., Liu, Y., Pajerowski, D., Yan, J., Bridges, C. A., Savici, A. T. (2018), \emph{npj Quantum Materials}, \textbf{3}, 8.} 
\reference{Basnet, R., Kotur, K. M., Rybak, M., Stephenson, C., Bishop, S., Autieri, C., Birowska, M., Hu, J. (2022), \emph{Phys. Rev. Res.} \textbf{4}, 023256.} 
\reference{Bera, A. K., Yusuf, S. M., Kumar, A, Ritter, C. (2017), \emph{Phys. Rev. B} \textbf{95}, 094424.} 
\reference{Bera, A. K., Yusuf, S. M., Keller, L., Yokaichiya, F., Stewart, J. R. (2022), \emph{Phys. Rev. B} \textbf{105}, 014410.} 
\reference{Bera, A. K., Yusuf, S. M., Orlandi, F.,  Manuel, P., Bhaskaran, L., Zvyagin, S. A. (2023) \emph{Phys. Rev. B} \textbf{108}, 214419} 
\reference{Berthelot, R., Schmidt, W., Muir, S., Eilertsen, J., Etienne, L., Sleight, A. W., Subramanian, M. A. (2012), \emph{Inorg. Chem.} \textbf{51}, 5377–5385.}
\reference{Biffin, A., Johnson, R. D., Choi, S., Freund, F., Manni, S.,  Bombardi, A.,  Manuel,bP.,  Gegenwart,P., Coldea, R. (2014), \emph{Phys. Rev. B} \textbf{90}, 205116.} 
\reference{Bose, A., Routh, M., Voleti, S., Saha, S.K., Kumar, M., Saha-Dasgupta, T. and Paramekanti, A. (2023), \emph{Physical Review B}, \textbf{108}, 174422.} 
\reference{Bramwell, S. T.,  Buckley, A. M., Day, P. (1994) \emph{J. Solid State Chem.} \textbf{111}, 48.} 
\reference{Bravyi, S.B. and Kitaev, A.Y., (2002), \emph{Annals of Physics}, \textbf{298}, 210-226.} 
\reference{Broholm, C., Cava, R.J., Kivelson, S.A., Nocera, D.G., Norman, M.R., and Senthil, T. (2020), \emph{Science} \textbf{367}, eaay0668.} 
\reference{Buyers, W.J.L., Holden, T.M., Svensson, E.C., Cowley, R.A. and Hutchings, M.T., (1971), \emph{J. Phys. C: Solid State Physics}, \textbf{4}, 2139.} 
\reference{Cao, H. B., Banerjee, A., Yan, J.-Q., \textit{et al} (2016), \emph{Phys. Rev. B} \textbf{93}, 134423.}  
\reference{Chakraborty, A., Kumar, V., Bachhar, S., B\"{u}ttgen, N., Yokoyama, K., Biswas, P. K., Siruguri, V., Pujari, S., Dasgupta, I. \& Mahajan, A. V. (2021), \emph{Phys. Rev. B} \textbf{104}, 115106.} 
\reference{Chen, W., Li, X., Hu, Z., Hu, Z., Yue, L., Sutarto, R., He, F., Iida, K., Kamazawa, K., Yu, W. and Lin, X., (2021),  \emph{Phys. Rev. B}, \textbf{103}, L180404.} 
\reference{Cheryl, W. \& Avdeev, M. \& Ling, C. D. (2016), \emph{J. Solid State Chem.} \textbf{243}, 18-22.} 
\reference{Das, S., Voleti, S., Saha-Dasgupta, T. and Paramekanti, A. (2021), \emph{Phys. Rev. B}, \textbf{104}, 134425.} 
\reference{Do, S. H., Park, S. Y., Yoshitake, J. \emph{et al.} (2017), \emph{Nat. Phys.} \textbf{13}, 1079.}
\reference{Dufault, E., Bahrami, F., Streeter, A., Yao, X., Gonzalez, E., Zhang, Q., Tafti, F. (2023), \emph{Phys. Rev. B} \textbf{108}, 064405.} 
\reference{Fouet, J.B., Sindzingre, P. and Lhuillier, C., (2001), \emph{The European Physical Journal B-Condensed Matter and Complex Systems} \textbf{20}, 241-254.} 
\reference{Freedman, M.H., Kitaev, A. and Wang, Z., (2002), \emph{Communications in Mathematical Physics}, \textbf{227}, 587-603.} 
\reference{Gao, B., Chen, T., Wang, C., Chen, L., Zhong, R., Abernathy, D. L., Xiao, D., Dai, P. (2021), \emph{Phys. Rev. B} \textbf{104}, 214432.}
\reference{Gass, S., Cônsoli, P.M., Kocsis, V., Corredor, L.T., Lampen-Kelley, P., Mandrus, D.G., Nagler, S.E., Janssen, L., Vojta, M., Büchner, B. and Wolter, A.U.B., (2020), \emph{Phys. Rev. B}, \textbf{101(24)}, 245158.} 
\reference{Goff, J.P., Tennant, D.A. and Nagler, S.E., (1995), \emph{Physical Review B}, \textbf{52}, 15992.} 
\reference{Gu, Y, Li, X., Gu, Y., Chen,Y., Iida, K., Nakao, A., Munakata, K, Garlea, V. O., Li,Y., Deng, G., Zaliznyak, I. A., Tranquada, J. M., Li, Yuan (2024), \emph{Phys. Rev. B} \textbf{109}, L060410.} 
\reference{Halloran, T., Desrochers, F., Zhang, E. Z., \textit{et al.} (20-22)\emph{Proc. Natl. Acad. Sci. USA} \textbf{120}, e2215509119.}
\reference{Hermanns, M., Kimchi, I., Knolle, J. (2018), \emph{Annual Review of Condensed Matter Physics}, \textbf{9}, 17-33.} 
\reference{Holden, T.M., Buyers, W.J.L., Svensson, E.C., Cowley, R.A., Hutchings, M.T., Hukin, D., Stevenson, R.W.H. (1971), \emph{J. Phy. C: Solid State Phys.}, \textbf{4(14)}, 2127.} 
\reference{Imai, Y., Kazuhiro N., Yasuhiro S., \textit{et al.} (2022), \emph{Phys. Rev. B} \textbf{105}, L041112.} 
\reference{Jackeli, G. \& Khaliullin, G. (2009), \emph{Phys. Rev. Lett.} \textbf{102}, 017205.}
\reference{Jiang, S., White, S.R. and Chernyshev, A.L., (2023), \emph{Physical Review B}, \textbf{108}, L180406.} 
\reference{Johnson, R. D., Williams, S. C., Haghighirad, A. A., Singleton, J., Zapf, V., Manuel, P., Mazin, I. I., Li, Y., Jeschke, H. O., Valent\'{\i}, R., Coldea, R. (2015), \emph{Phys. Rev. B} \textbf{92}, 235119.} 
\reference{Kanyolo, G. M., Masese, T., Alshehabi, A., Huang, Z.-D. (2023) \emph{Mater. Today Chem.} \textbf{33} 101657.}
\reference{Kasahara, Y., Ohnishi, T., Mizukami, Y., Tanaka, O., Ma, S., Sugii, K., Kurita, N., Tanaka, H., Nasu, J., Motome, Y. and Shibauchi, T., (2018), \emph{Nature}, \textbf{559}, 227-231.} 
\reference{Kenney, E. M., Segre,C. U., Lafargue-Dit-Hauret, W., Lebedev, O. I., Abramchuk, M., Berlie, A., Cottrell, S. P., Simutis, G., Bahrami, F., Mordvinova, N. E., Fabbris, G., McChesney, J. L., Haskel, D., Rocquefelte, X., Graf, M. J., Tafti, F. (2019) \emph{Phys. Rev. B} \textbf{100}, 094418.} 
\reference{Khuntia, P., Manni, S., Foronda, F. R., Lancaster, T., Blundell, S. J., Gegenwart, P., Baenitz, M. (2017), \emph{Phys. Rev. B} \textbf{96}, 094432.} 
\reference{Kim, C., Jeong, J., Park, P., Masuda, T., Asai, S., Itoh, S., Kim, H.S., Wildes, A. and Park, J.G., (2020), \emph{Physical Review B}, \textbf{102}, 184429.} 
\reference{Kimchi I., You, Y.-Z. (2011), \emph{Phys. Rev. B} \textbf{84}, 180407(R).}
\reference{Kitaev, A. Y., (2003), \emph{Annals of Physics}, \textbf{303}, 2-30.} 
\reference{Kitaev, A., (2006), \emph{Annals of Physics}, \textbf{321}, 2-111.} 
\reference{Kitagawa, K., Takayama, T., Matsumoto, Y. \emph{et al.} (2018), \emph{Nature} \textbf{554}, 341–345.} 
\reference{Korshunov, S. E., (1993), \emph{Phys. Rev. B}, \textbf{47}, 6165(R).} 
\reference{Kr\"uger, W. G. F., Chen, W., Jin, X., Li, Y., Janssen, L., (2023) \emph{Phys. Rev. Lett.} \textbf{131}, 146702 }
\reference{Kubota, Y., Tanaka, H., Ono, T., Narumi, Y. and Kindo, K., (2015), \emph{Physical Review B}, \textbf{91}, 094422.} 
\reference{Kurbakov, A. I., Korshunov, A. N., Podchezertsev, S. Yu., Malyshev, A. L., Evstigneeva, M. A., Damay, F., Park, J., Koo, C., Klingeler, R., Zvereva,E. A, Nalbandyan, V. B (2017), \emph{Phys. Rev. B} \textbf{96}, 024417.} 
\reference{Kurbakov, A. I., Korshunov, A. N., Podchezertsev, S. Yu, Stratan, M. I., Raganyan, G. V., Zvereva, E. A. (2020), \emph{J. Alloys Comp.} \textbf{820}, 153354.} 
\reference{Lampen-Kelley, P., Rachel, S., Reuther, J., Yan, J. Q., Banerjee, A., Bridges, C.A., Cao, H.B., Nagler, S.E. and Mandrus, D., (2018), \emph{Physical Review B}, \textbf{98(10)}, 100403.} 
\reference{Lefran\c{c}ois, E., Songvilay, M., Robert, J., Nataf, G., Jordan, E., Chaix, L., Colin, C. V., Lejay, P., Hadj-Azzem, A., Ballou, R., Simonet, V. (2016), \emph{Phys. Rev. B} \textbf{94}, 214416.}
\reference{Li, P. H. Y., Bishop, R. F. and Campbell, C. E. (2014), \emph{Phys. Rev. B}, \textbf{89}, 220408.} 
\reference{Li, X., Gu, Y., Chen,Y., Garlea, V. O., Iida, K., Kamazawa, K. Li,Y., Deng, G.,  Xiao, Q., Zheng, X., Ye, Z., Peng, Y., Zaliznyak, I. A., Tranquada, J. M., Li, Yuan, (2022), \emph{Phys. Rev. X}, \textbf{12}, 041024.} 
\reference{Liu, H. and Khaliullin, G., (2018), \emph{Phys. Rev. B}, \textbf{97}, 014407.} 
\reference{Liu, H., Chaloupka, J. and Khaliullin, G. (2020), \emph{Phys. Rev. Lett.}, \textbf{125}, 047201.} 
\reference{Liu, X. and Kee, H. Y., (2023) \emph{Phys. Rev. B}, \textbf{107}, 054420.} 
\reference{Little, A., Wu, L., Lampen-Kelley, P., Banerjee, A., Patankar, S., Rees, D., Bridges, C.A., Yan, J.Q., Mandrus, D., Nagler, S.E. and Orenstein, J. (2017), \emph{Phys. Rev. Lett.}, \textbf{119}, 227201.} 
\reference{Maksimov, P. A., Ushakov, A. V., Pchelkina, Z. V., Li, Y., Winter, S. M. and Streltsov, S. V., (2022), \emph{Phys. Rev. B}, \textbf{106}, 165131.} 
\reference{Matsubara, N., Nocerino, E., Forslund, O.K. \emph{et al.} (2020) \emph{Sci Rep} \textbf{10}, 18305.} 
\reference{Moessner, R., Sondhi, S.L. and Chandra, P., (2001), \emph{Phys. Rev. B}, \textbf{64}, 144416.} 
\reference{Momma, K., Izumi, F. (2011) \emph{J. Appl. Cryst.}, \textbf{44}, 1272}
\reference{Morgan, Z., \emph{et al.} (2024), \emph{Phys. Rev. Mat.} \textbf{8}, 016201.}
\reference{Motome, Y., Sano, R., Jang, S., Sugita, Y. and Kato, Y. (2020), \emph{J. Phys.: Cond. Matt.}, \textbf{32}, 404001.} 
\reference{Mu, S., Dixit, K. D., Wang, X., Abernathy, D. L., Cao, H., Nagler, S.E., Yan, J., Lampen-Kelley, P., Mandrus, D., Polanco, C.A. and Liang, L. (2022), \emph{Phys. Rev. Res.}, \textbf{4}, 013067.} 
\reference{Park, Sang-Youn \textit{et al} (2024) \emph{J. Phys.: Cond. Matt.} \textbf{36} 215803.}
\reference{Perez-Mato, J. M., Gallego, S. V., Tasci, E., Elcoro, L., de la Flor, G. \& Aroyo, M. I. (2015), \emph{Annu. Rev. Mater. Res.} \textbf{45}, 217.} 
\reference{Pressley, L., \textit{et al.} (2024), \emph{in preparation} }
\reference{Price, C. and Perkins, N. B. (2013), \emph{Phys. Rev. B}, \textbf{88(2)}, 024410.} 
\reference{Qi, C., Jiang, F. and Yang, S. (2021), \emph{Composites Part B: Engineering}, \textbf{227}, 109393.} 
\reference{Rau, J. G., McClarty, P. A. and Moessner, R. (2018), \emph{Phys. Rev. Lett.}, \textbf{121(23)}, 237201.} 
\reference{Regnault, L.P., Burlet, P., Rossat-Mignod, J. (1977), \emph{Physica B+C}, \textbf{ 86–88}, 660-662.}
\reference{Regnault,L. P., Henry, J. Y., Rossat-Mignod, J., De Combarieu, A. (1980), \emph{J. Magn. Magn. Mater} \textbf{15-18}, 1021.}
\reference{Regnault, L., Boullier, C., Lorenzo, J.E. (2018). \emph{Heliyon}, \textbf{4}.} 
\reference{Ringler, J.A., Kolesnikov, A.I. and Ross, K.A. (2022). \emph{Phys. Rev. B}, \textbf{105}, 224421.} 
\reference{Rodríguez-Carvajal, J. (1993), \emph{Phys. B}, \textbf{192}, 55.}
\reference{Roudebush, J. H., Andersen, N. H., Ramlau, R., Garlea, V. O., Toft-Petersen, R., Norby, P., Schneider, R., Hay, J. N., Cava R. J. (2013), \emph{Inorg. chem.} \textbf{52}, 6083-6095.} 
\reference{Samarakoon, A. M., Chen, Q., Zhou, H., Garlea, V. O. (2021), \emph{Phys. Rev. B} \textbf{104}, 184415.} 
\reference{Sandilands, L.J., Tian, Y., Plumb, K.W., Kim, Y.J. and Burch, K.S., (2015), \emph{Physical review letters}, \textbf{114(14)}, 147201.} 
\reference{Savary, L. and Balents, L. (2016), \emph{Rep. Prog. Phys.} \textbf{80}, 016502.} 
\reference{Sears, J. A., Songvilay, M., Plumb, K. W., Clancy, J. P., Qiu, Y., Zhao, Y., Parshall, D., Kim, Y.-J. (2015), \emph{Phys. Rev. B} \textbf{91}, 144420.} 
\reference{Sears, J. A., Chern, L. E., Kim, S., Bereciartua, P. J., Francoual, S., Kim, Y. B. and Kim, Y. J., (2020), \emph{Nature physics}, \textbf{16}, 837.} 
\reference{Seepersad, C. C., Dempsey, B. M., Allen, J. K., Mistree, F., McDowell, D.L. (2004), \emph{AIAA journal}, \textbf{42}, 1025.} 
\reference{Seibel, E. M., Roudebush, J. H., Wu, H., Huang, Q., Ali, M. N., Ji, H. \& Cava, R. J. (2013), \emph{Inorg. Chem.} \textbf{52}, 13605.} 
\reference{Scheie, A., Ross, K., Stavropoulos, P. P, Seibel, E., Rodriguez-Rivera, J. A., Tang, J. A., Li, Yi, Kee, H.-Y, Cava, R. J., Broholm, C. (2019), \emph{Phys. Rev. B} \textbf{100}, 214421.} 
\reference{Scheie, A., Park, P., Villanova, J.W., Granroth, G.E., Sarkis, C.L., Zhang, H., Stone, M.B., Park, J.G., Okamoto, S., Berlijn, T. and Tennant, D.A., (2023), \emph{Phys. Rev. B}, \textbf{108}, 104402.} 
\reference{Shangguan, Y., Bao, S., Dong, ZY. \emph{et al.} (2023)\emph{Nat. Phys}. \textbf{19}, 1883.} 
\reference{Sindzingre, P., Lhuillier, C., Fouet, J. B. (2003), \emph{International Journal of Modern Physics B}, \textbf{17}, 5031-5039.} 
\reference{Songvilay, M,. Robert, J., Petit, S., Rodriguez-Rivera, J. A., Ratcliff, W. D., Damay, F., Bal\'{e}dent, V., Jim\'{e}nez-Ruiz, M., Lejay, P., Pachoud, E., Hadj-Azzem, A., Simonet, V., Stock, C. (2020) \emph{Phys. Rev. B} \textbf{102}, 224429.}
\reference{Stavropoulos, P. P., Pereira, D. \& Kee, H.-Y. (2019), \emph{Phys. Rev. Lett.} \textbf{123}, 037203.}
\reference{Stratan, M. I., Shukaev, I. L., Vasilchikova, T. M., Vasiliev, A. N., Korshunov, A. N., Kurbakov, A. I., Nalbandyan, V. B., Zvereva, E. A. (2019), \emph{New J. Chem.} \textbf{43}, 13545-13553.} 
\reference{Taddei, K. M., Garlea, V. O., Samarakoon, A. M., Sanjeewa, L. D., Xing, J., Heitmann, T. W., dela Cruz, C., Sefat, A. S., Parker, D. (2023), \emph{Phys. Rev. Res.} \textbf{5},  013022.} 
\reference{Takagi, H., Takayama, T., Jackeli, G., Khaliullin, G. and Nagler, S. E. (2019), \emph{Nature Reviews Physics}, \textbf{1}, 264} 
\reference{Trebst, S. and Hickey, C. (2022), \emph{Phys. Reports} \textbf{950}, 1-37.} 
\reference{Vasilchikova, T., Vasiliev, A., Evstigneeva, M., Nalbandyan, V., Lee, J.-S., Koo, H.-J., Whangbo, M.-H. (2022), Materials \textbf{15}, 2563.} 
\reference{Vivanco, H. K. \& Trump, B. A. \& Brown, C. M. \& McQueen, T. M. (2020), \emph{Phys. Rev. B} \textbf{102}, 224411.} 
\reference{Wang, J. and Liu, Z. X. (2023), \emph{Phys. Rev. B} \textbf{108}, 014437.} 
\reference{Werner, J., Hergett, W., Park, J., Koo, C., Zvereva, E. A., Vasiliev, A. N., Klingeler R.  (2019), \emph{J. Magn. Magn. Mat.} \textbf{481}, 100.} 
\reference{Wildes, A. R., Simonet, V., Ressouche, E., McIntyre, G. J., Avdeev, M., Suard, E., Kimber, S. A. J., Lan\c{c}on, D., Pepe, G., Moubaraki, B., Hicks, T. J. (2015), \emph{Phys. Rev. B} \textbf{92}, 224408.} 
\reference{Wildes, A. R., Simonet, V., Ressouche, E., Ballou, R., McIntyre, G. J. (2017), \emph{J. Phys.: Cond. Matt.} \textbf{29}, 455801.} 
\reference{Wildes, A. R., Stewart, J. R., Le, M. D., Ewings, R. A., Rule, K. C., Deng, G., Anand, K. (2022), \emph{Phys. Rev. B} \textbf{106}, 174422.}
\reference{Wildes, A. R., F{\aa}k, B., Hansen, U. B., Enderle, M., Stewart, J. R., Testa, L., R{\o}nnow, H. M., Kim, C., Park, Je-Geun (2023), \emph{Phys. Rev. B} \textbf{107}, 054438.}
\reference{Williams, S. C., Johnson, R. D., Freund, F., Choi, S., Jesche, A., Kimchi, I., Manni, S., Bombardi, A., Manuel, P., Gegenwart, P., Coldea, R. (2016), \emph{Phys. Rev. B} \textbf{93}, 195158.} 
\reference{Winter, S. M., Tsirlin, A.A., Daghofer, M., van den Brink, J., Singh, Y., Gegenwart, P. and Valentí, R., (2017), \emph{Journal of Physics: Condensed Matter}, \textbf{29}, 493002.} 
\reference{Winter, S.M. (2022), \emph{J. Phys.: Materials}, \textbf{5}, 045003.} 
\reference{Wong, C., Avdeev, M., Ling Ch. D. (2016) \emph{J. Solid State Chem.},\textbf{143},18} 
\reference{Xu, X., Wu, L., Zhu, Y., Cava, R. J. (2023), \emph{Phys. Rev. B} \textbf{108}, 174432.} 
\reference{Yadav, R., Nishimoto, S., Richter, M., van den Brink, J., Ray, R. (2019), \emph{Phys. Rev B} \textbf{100}, 144422 } 
\reference{Yan, J-Q., Okamoto, S., Wu, Y., Zheng, Q., Zhou, H. D., Cao, H. B., McGuire, M. A. (2019), \emph{Phys. Rev. Mat.} \textbf{3}, 074405.} 
\reference{Yao, W., Iida, K., Kamazawa, K., Li, Yuan {2022} \emph{Phys. Rev. Lett.} \textbf{129} 147202.}
\reference{Yao, W., Zhao, Y., Qiu, Y., Balz, Ch., Stewart, J. R., Lynn, J. W., Li, Yuan (2023), \emph{Phys. Rev. Res} \textbf{5}, L022045} 
\reference{Ye, F., Chi, S., Cao, H., Chakoumakos, B. C., Fernandez-Baca, J. A., Custelcean, R., Qi, T. F., Korneta, O. B., Cao, G (2012), \emph{Phys. Rev. B} \textbf{85}, 180403.} 
\reference{Yokoi, T., Ma, S., Kasahara, Y., Kasahara, S., Shibauchi, T., Kurita, N., Tanaka, H., Nasu, J., Motome, Y., Hickey, C. and Trebst, S., (2021), \emph{Science}, \textbf{373(6554)}, 568-572.} 
\reference{Zhang, H. and Lamas, C. A. (2013),  \emph{Phys. Rev. B}, \textbf{87}, 024415.} 
\reference{Zhang, Q., Yang, X., Li, P., Huang, G., Feng, S., Shen, C., Han, B., Zhang, X., Jin, F., Xu, F., Lu, T.J. (2015), \emph{Prog. Mater. Sci.}, \textbf{74}, 332.} 
\reference{Zhang, Sh., Lee, S., Woods, A. J., Peria, W. K., Thomas, S. M., Movshovich, R., Brosha, E., Huang, Q., Zhou, H., Zapf, V. S., Lee, M. (2023), \emph{Phys. Rev. B} \textbf{108}, 064421.} 
\reference{Zvereva, E. A., Stratan, M. I., Ovchenkov, Y. A., Nalbandyan, V. B., Lin, J.-Y., Vavilova, E. L., Iakovleva, M. F., Abdel-Hafiez, M., Silhanek, A. V., Chen, X.-J., Stroppa, A., Picozzi, S., Jeschke, H. O., Valent\'{\i}, R., Vasiliev, A. N. (2015), \emph{Phys. Rev. B} \textbf{92}, 144401.}
\reference{Zvereva, E. A., Stratan, M. I.,  Ushakov, A. V., Nalbandyan, V. B., Shukaev, I. L., Silhanek, A. V., Abdel-Hafiez, M., Streltsov, S. V., Vasiliev, A. N. (2016), \emph{Dalton Trans.} \textbf{45}, 7373.} 

\end{references}
\end{document}